\documentclass[utf8]{frontiersFPHY} 

\usepackage{url,hyperref,lineno,microtype,subcaption}
\usepackage[onehalfspacing]{setspace}
\usepackage{multirow}

\usepackage{aas_macros}

\def\keyFont{\fontsize{8}{11}\helveticabold }
\def\firstAuthorLast{Yu {et~al.}} 
\def\Authors{Hang Yu\,$^{1,*}$ and Rana X. Adhikari\,$^{2}$}
\def\Address{$^{1}$TAPIR, Walter Burke Institute for Theoretical Physics, MC 350-17 California Institute of Technology, Pasadena, CA 91125, USA\\
$^{2}$LIGO Laboratory, MC 100-36, California Institute of Technology, Pasadena, CA 91125, USA  }
\def\corrAuthor{Hang Yu}

\def\corrEmail{hangyu@caltech.edu}

\begin{document}
\onecolumn
\firstpage{1}

\title[Nonlinear noise regression]{Nonlinear noise regression in gravitational-wave detectors with convolutional neural networks} 

\author[\firstAuthorLast ]{\Authors} 
\address{} 
\correspondance{} 

\extraAuth{}

\maketitle

\begin{abstract}

\section{}
Currently, the sub-60 Hz sensitivity of gravitational-wave (GW) detectors like Advanced LIGO is limited by the control noises from auxiliary degrees of freedom, which nonlinearly couple to the main GW readout. One particularly promising way to tackle this contamination is to perform nonlinear noise mitigation using machine-learning-based convolutional neural networks (CNNs), which we examine in detail in this study. As in many cases the noise coupling is bilinear and can be viewed as a few fast channels' outputs modulated by some slow channels, we show that we can utilize this knowledge of the physical system and adopt an explicit ``slow$\times$fast'' structure in the design of the CNN to enhance its performance of noise subtraction. We then examine the requirement in the signal-to-noise ratio (SNR) in both the target channel (i.e., the main GW readout) and in the auxiliary sensors in order to reduce the noise by at least a factor of a few. In the case of limited SNR in the target channel, we further demonstrate that the CNN can still reach a good performance if we adopt curriculum learning techniques, which in reality can be achieved by combining data from quiet times and those from periods with active noise injections.

\tiny
 \keyFont{ \section{Keywords:} gravitational-wave detectors, Advanced LIGO, noise regression, machine learning, neural networks} 
\end{abstract}




\section{Introduction}

Since September 14, 2015~\citep{Abbott:2016blz}, gravitational-wave (GW) observatories including Advanced LIGO (aLIGO; \citealt{LSC:15}), Advanced Virgo~\citep{TheVirgo:2014hva}, and KAGRA~\citep{Akutsu:2018axf} have achieved a great success with dozens of GW event detected so far~\citep{LIGOScientific:2018mvr, Abbott:2020niy}. 

While the high-frequency ($\gtrsim 60\,{\rm Hz}$) part of aLIGO's sensitivity is steadily approaching its designed target especially with the implementation of quantum squeezing \citep{Tse:19}, there is nonetheless a big gap between the current and the designed sensitivity at lower frequencies \citep{Martynov:16, Buikema:20}. 

If we can remove the excess contamination in the sub-60 Hz band, it would greatly benefit a large array of science including the early warning of binary neutron star mergers \citep{Cannon:12,LSC:19,Sachdev:20,Chu:20,Yu:21c}, the detection of intermediate-mass black holes \citep{Mandel:08, Graff:15, Veitch:15, GW190521}, the constraining of eccentricities of binary black holes and hence their formation channels~\citep{Romero-Shaw:19, LSC:19b}, and many more. See also the discussions in, e.g., \citet{Yu:18} and references therein. 

What limit the current sensitivity below 60 Hz are ``technical noises'' due to environmental perturbations and the control noises of auxiliary degrees of freedom. Unlike fundamental noises due to quantum and thermal fluctuations in the main GW readout channel that cannot be mitigated without instrumental upgrades, the technical noises can in principle be removed via regression techniques as their source fluctuations are also witnessed and recorded by hundreds of auxiliary sensors (i.e., sensors that do not detect GW signals) employed by aLIGO. In fact, the linear component of technical noises has successfully been removed in aLIGO (see, e.g., \citealt{Driggers:19, Davis:19}). 

The remaining challenge of the regression problem is to tackle noises that couple to the main GW readout \emph{nonlinearly}. In fact, many noise sources in aLIGO couple in a bilinear way to the GW readout. It happens naturally when the coupling coefficient of an auxiliary channel is modulated by some slow motion ($\lesssim 1\,{\rm Hz}$) in the interferometer. The modulation destroys the linear coherence between the auxiliary channel and the GW readout, forbidding the use of standard linear regression technique like the one employed in \citet{Driggers:19}. Moreover, it is typically challenging to reconstruct the modulation directly because of the pollution from large ambient motion as well as cross-couplings from complicated control feedback loops stabilizing aLIGO at below 1 Hz. 

Fortunately, machine-learning (ML) techniques, especially the use of convolutional neural networks (CNNs), offers an attractive potential solution to the nonlinear noise regression problem (see, e.g., \citealt{ Ormiston:20, Vajente:20, Mogushi:21, Yu:21c}). By inputting to a CNN sufficiently many auxiliary witnesses that contain all the information about the noise coupling, and utilizing properly designed network structure and training strategy, we can let the machine to figure out the coupling mechanism behind even it involves nonlinearity and complicated blending of different sensors. Furthermore, after the training process, the subsequent predicting of the contamination using CNN is highly efficient computationally, allowing the regression to be performed in real time (i.e., online). This would be especially beneficial for searches that requires low latency, such as the early warning of binary neutron star mergers \citep{Baltus:21, Yu:21c}. Other successful useage of ML techniques in GW astronomy including the identification of various GW events~\citep{Huerta:20, Krastev:20, Chan:20, Dreissigacker:20, Wong:20, Schafer:20, Bayley:20, Wei:21c, Chang:21, Yan:21, SaizPerez:21, Chatterjee:21, Mishra:21, Lopez:21, Beheshtipour:21, Marianer:21}, source parameter estimations~\citep{Gabbard:19, Chua:20, Green:20, Talbot:20, Chatterjee:20, DEmilio:21, Alvares:21,  Williams:21, Krastev:21, Xia:21}, and detector characterization~\citep{Colgan:20, Essick:20, Cuoco:20, Torres-Forne:20, Biswas:20, Soni:21, Sankarapandian:21, Zhan:21, Mogushi:21b}.   

In this work, we thus explore in detail how we could use ML-based CNN to potentially mitigate the angular noise in aLIGO, which is the limiting noise source of the current sensitivity in the $30\,{\rm Hz}$ band and is also a classical example of bilinear coupling in aLIGO. Our implementation utilizes the code \texttt{Keras} \citep{Chollet:15}, a deep learning application programming interface written in \texttt{Python}, running on top of the ML platform \texttt{TensorFlow} \citep{Abadi:15}.

The rest of the paper is organized as follows. In Sec.~\ref{sec:sim_noise}, we describe the coupling mechanism behind the angular noise and how we generate mock data so that our study can be carried out in a controlled way. We then explore in Sec.~\ref{sec:NN_structure} how does the structure of a CNN affect its performance, with a focus on the comparison between a general structure and one inspired by the physics. In Sec.~\ref{sec:snr_req}, we assess the CNN performance as a function of the signal-to-noise ratio, or SNR, in both the target channel (GW readout) and in the input witnesses. This is followed by Sec.~\ref{sec:CL} where we demonstrate the use of curriculum learning may help the convergence of a CNN when its SNR is low in the target. Lastly, we conclude and discuss in Sec.~\ref{sec:summary}.


\section{Simulating noises in the LIGO system}
\label{sec:sim_noise}

\begin{figure}[htb]
\begin{center}
\includegraphics[width=0.5\textwidth]{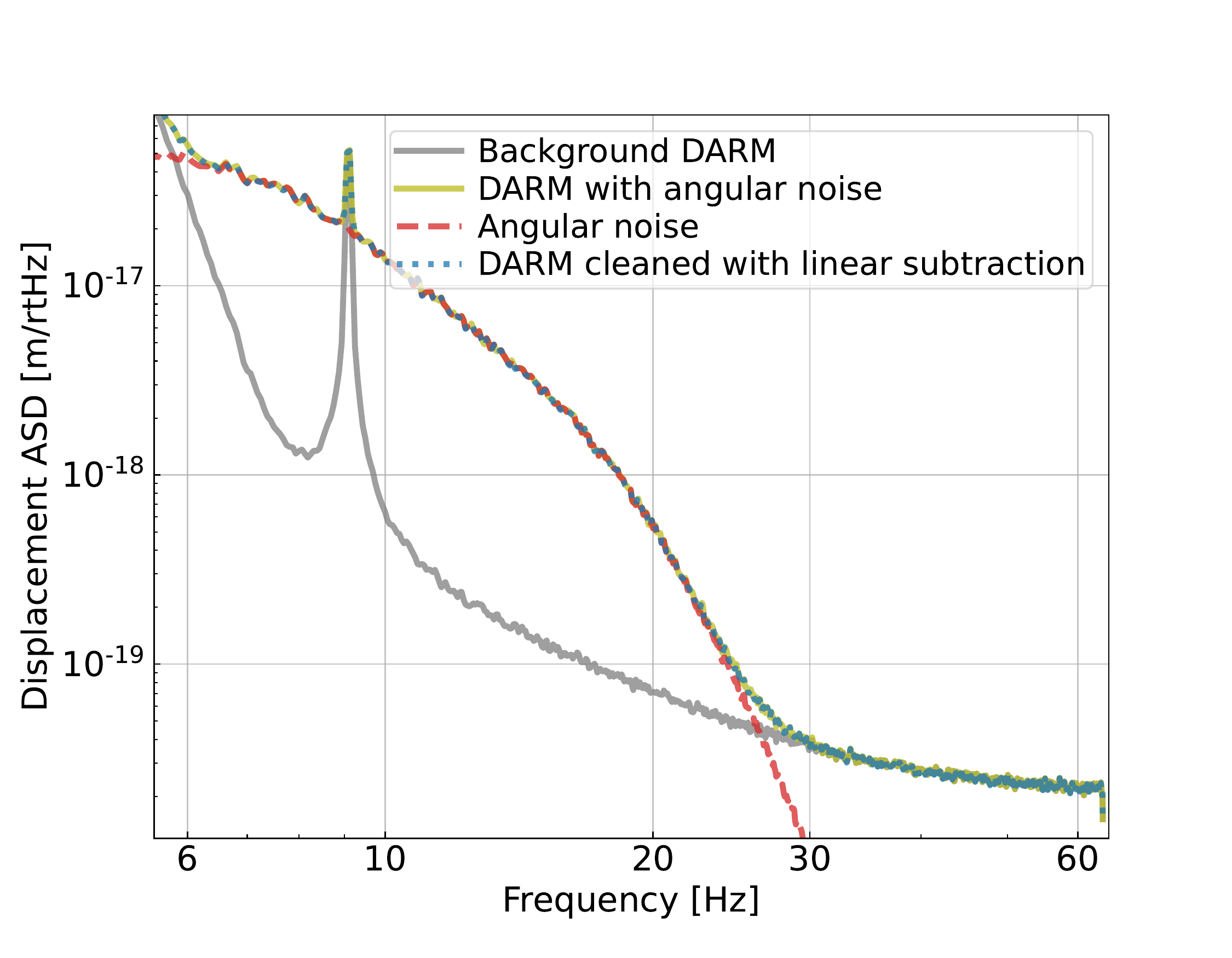}
\end{center}
\caption{Simulated ASD of the GW readout (``DARM'') including both a fundamental component (grey trace given by the designed sensitivity; \citealt{LSC:15}) and an extra contamination due to the angular noise [Eq.~(\ref{eq:bilin_cav})]. Because the coupling mechanism is bilinear, a standard linear subtraction cannot mitigate the contamination as shown in the dotted-blue trace.  }\label{fig:bilin_darm_asd}
\end{figure}

In this work, we will focus on using ML-based CNN to mitigate noise due to the angular control system in LIGO, which is one of the noise sources limiting the sub-30 Hz sensitivity of LIGO currently~\citep{Buikema:20}. To make our study controlled, we will use simulated time series for various channels with characteristics similar to the real aLIGO system. Throughout this study, we will use a fixed sampling rate of 128\,Hz for all the time series.

In Fig.~\ref{fig:bilin_darm_asd} we show a typical plot of the amplitude spectral density (ASD) of our simulated GW readout including both the fundamental noise (gray trace; which cannot be further reduced by offline regression) and the simulated angular noise (red-dashed trace). Note that the coupling of the angular noise to the main GW readout is bilinear (which we will describe in detail shortly). Consequently, a standard linear regression fails to mitigate its contamination (blue-dotted trace), which thus motivates us to investigate regression strategies utilizing ML CNN. 
In this Section, we present details of how we simulate the angular noise (red trace in Fig.~\ref{fig:bilin_darm_asd}). 
The subtraction of the simulated noise will then be presented in Secs.~\ref{sec:NN_structure}-\ref{sec:CL}.

At the power level LIGO is currently operating at, the angular control noise couples to the main GW readout (i.e., the monitor of Differential ARM length, or ``DARM'') via a geometrical effect. If the beam spot is not at the center of the rotational pivot of the mirror, then an angular motion will be converted to a length fluctuation bilinearly as~\citep{Barsotti:10, Yu:19}
\begin{equation}
    \delta x^{\rm (mir)}(t) = x_{\rm spot}^{\rm (mir)} (t) \theta^{\rm (mir)}(t).
    \label{eq:bilin_single_mirror}
\end{equation}
Here $\theta^{\rm (mir)}(t)$ corresponds a fast ($\gtrsim 10\,{\rm Hz}$) angular perturbation to the mirror, which is further induced by the sensing noise in the angular control system being fed back to the mirror. The $x_{\rm spot}^{\rm (mir)}$, on the other hand, corresponds to a slow ($\lesssim 1\,{\rm Hz}$) motion of the beam spot on the test mass induced by the seismic motion.

Note that the Eq.~(\ref{eq:bilin_single_mirror}) describes the local length fluctuation of a single mirror [and hence a superscript ``(mir)'' is added to each quantity]. What of physical interest is in fact the length fluctuation of an ARM cavity form by an input test mass and an end test mass. I.e., we can measure the relative distance between the two test masses (which are also mirrors), and the GW signal is further readout via the differential-ARM (DARM) length.
Therefore, the angular noise affects the cavity length as
\begin{equation}
    \delta x(t) = 
    \begin{bmatrix}
        x^{\rm (e)}_{\rm spot}(t), & x^{\rm (i)}_{\rm spot} (t)
    \end{bmatrix}
    \begin{bmatrix}
        \theta^{\rm (e)}(t) \\ 
        \theta^{\rm (i)}(t) 
    \end{bmatrix},
    \label{eq:bilin_cav}
\end{equation}
where the superscript ``(e)'' [``(i)''] stands for the end (input) test mass. Its contamination to DARM is further obtained by first incoherently simulating each cavity's length according to Eq.~(\ref{eq:bilin_cav}) and then take the difference. 

In reality, the mirrors are not controlled locally (in the ``mirror basis'') but instead in a radiation-torque basis. This choice is to tackle the Sidles-Sigg effect \citep{Sidles:06, Hirose:10, Dooley:13}: as the alignment changes, it creates a radiation torque feeding back to the alignment. In other words, the alignments of the input and end test masses are no more independent but coupled together via the radiation torque. Depending on the sign of the feedback, the radiation torque either hardens (i.e., makes it stiffer) or softens (makes it less stiff) the restoring torque of pendulums suspending the test masses. This allows us to decompose the alignments of the two test masses into a hard mode and a soft mode. The LIGO angular control is then performed in terms of this radiation pressure basis. 

Mathematically, we note that the radiation torque depends on the spot motions on the test masses, which further relate to the alignment of the test masses via the cavity geometry. For pitch motion (yaw motion is similar), this is given by~\citep{Sidles:06, Barsotti:10, Yu:19}
\begin{equation}
    \begin{bmatrix}
        x^{\rm (e)}_{\rm spot} \\ 
        x^{\rm (i)}_{\rm spot}
    \end{bmatrix}
    =\frac{L}{g^{\rm (e)}g^{\rm (i)}-1}
    \begin{bmatrix}
        g^{\rm (i)}, & 1 \\ 
        1, & g^{\rm (e)}
    \end{bmatrix}
    \begin{bmatrix}
        \theta^{\rm (e)} \\ 
        \theta^{\rm (i)} 
    \end{bmatrix}
    \equiv\boldsymbol{M}_{\rm a2s}
    \begin{bmatrix}
        \theta^{\rm (e)} \\ 
        \theta^{\rm (i)} 
    \end{bmatrix},
    \label{eq:a2s}
\end{equation}
where $L$ is the length of the ARM cavity, and $g^{\rm (e)}=1- L/R^{\rm (e)}$ [and similarly $g^{\rm (i)}=1- L/R^{\rm (i)}$] with $R^{\rm (e)}$ [$R^{\rm (i)}$] the radius of curvature of the end (input) test mass. For aLIGO, we have $L=3995\,{\rm m}$, $g^{\rm (e)}=-0.779$, and $g^{\rm (i)}=-1.065$~\citep{Barsotti:10}. The hard (soft) mode then corresponds to the eigenvector of the $\boldsymbol{M}_{\rm a2s}$ matrix associated with a positive (negative) eigenvalue. The conversion between the mirror basis ($\left[\theta^{\rm (e)}, \theta^{\rm (i)}\right]^T$) and the radiation-torque basis ($\left[\theta^{\rm (h)}, \theta^{\rm (s)}\right]^T$) is given by
\begin{equation}
    \begin{bmatrix}
        \theta^{\rm (h)} \\ 
        \theta^{\rm (s)} 
    \end{bmatrix}
    =\boldsymbol{M}_{\rm m2r}
    \begin{bmatrix}
        \theta^{\rm (e)} \\ 
        \theta^{\rm (i)} 
    \end{bmatrix}
    =
    \begin{bmatrix}
        0.756, & -0.655\\ 
        0.655, & 0.756
    \end{bmatrix}
    \begin{bmatrix}
        \theta^{\rm (e)} \\ 
        \theta^{\rm (i)} 
    \end{bmatrix},
    \label{eq:mirror_to_rad}
\end{equation}
where in the second equality we have plugged in numerical values for the aLIGO interferometer. 

In aLIGO, the tolerance on the residual hard-mode motion is more stringent and therefore it has a greater control bandwidth (unity-gain frequency ${\rm UGF}\simeq 3\,{\rm Hz}$) than the soft mode (${\rm UGF}\lesssim 1\,{\rm Hz}$). As a result, only the hard-mode control feeds back a significant amount of its sensing noise in the $10-30\,{\rm Hz}$ band while  the soft mode has negligible motion at high frequency. Consequently, in our simulation we only simulate $\theta^{\rm (h)}$ for the high-frequency angular motion and set $\theta^{\rm (s)}=0$. The typical sensing noise of the hard mode has a white amplitude spectral density (ASD) of $\sim 5\times 10^{-14}\,{\rm rad\,Hz^{-1/2}}$. Above the UGF, the physical $\theta^{(h)}$ is the sensing noise low-passed by the open-loop gain of the angular control loop and its spectral shape corresponds to the red trace in Fig.~\ref{fig:bilin_darm_asd} (see also, \citealt{Martynov:16, Buikema:20, Yu:19}). Once we have $\theta^{\rm (h)}$, it can then be converted back to the mirror basis by inverting Eq.~(\ref{eq:mirror_to_rad}).

\begin{figure}[htb]
\begin{center}
\includegraphics[width=\textwidth]{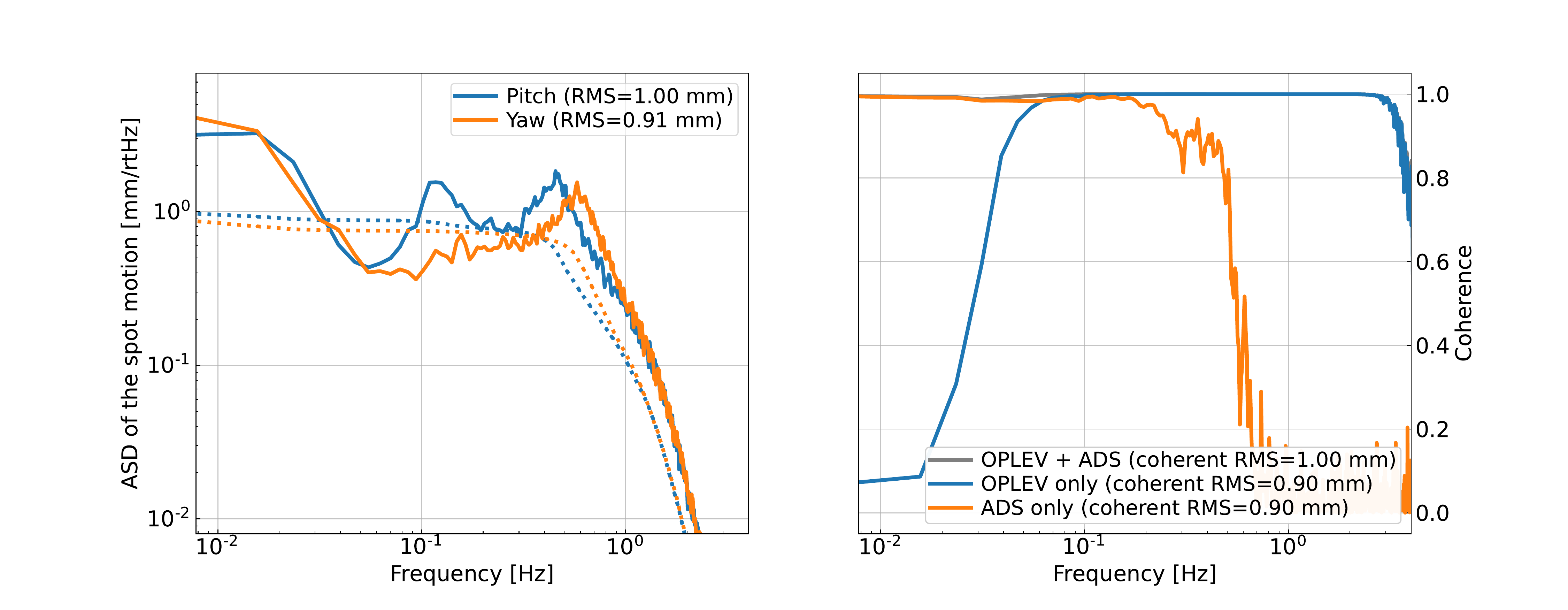}
\end{center}
\caption{Left panel: ASDs of the simulated beam-spot motions on one of the test masses. Right panel: the coherence between the true spot motion and the simulated witness channels. The $>0.1\,{\rm Hz}$ motion can be well sense by a pair of OPLEVs (for the input and end test masses), whose outputs are the angular motion relative to the local chamber. We assume that the $<0.1\,{\rm Hz}$ motion is sensed by an ADS channel which senses the spot by measuring the angle-to-length coupling coefficient. }\label{fig:bilin_spot}
\end{figure}

After introducing the high-frequency angular motion $\theta(t)$, we now describe how we simulate the low-frequency spot motion $x_{\rm spot}(t)$. Here we simulate directly the motion of each mirror.\footnote{Because the hard-mode and soft-mode control loops have different shape, the spot motion in the hard mode and in the soft mode in principle will have different values (similar to the case of high-frequency angular motion). However, subtleties arise because the spot motion governing the length fluctuation in Eq.~(\ref{eq:bilin_single_mirror}) is measured with respect to a local rotational pivot, whereas the angular control is performed relative to an input beam (i.e., a global quantity). At low-frequencies ($ <1\,{\rm Hz}$), the two reference points drift independently. Our current simulation does not capture this effect and for simplicity we treat the spot motion on each mirror as independent.} 
For an ideal suspension, the mirror should only have a significant amount of angular motion in pitch as at the suspension point, the seismic motion is mostly longitudinal and the suspension should only couple it into pitch but not yaw. However, in reality the suspension is not perfectly balanced; the imperfections in control loops also introduce cross-couplings between different degrees of freedom. As a result, the total root-mean-squared (RMS) motions in pitch and in yaw can be comparable. Therefore, in our simulation we assume they have similar RMS values of $\simeq 1\,{\rm mm}$. We nonetheless give them different spectral shapes as shown in the left panel of Fig.~\ref{fig:bilin_spot} as a challenge mimicking the real aLIGO system for our CNN to tackle. 

Unlike the high-frequency angular fluctuations that are directly readout in the control sensors, in aLIGO there are no sensors that directly probe the low-frequency spot motions over the entire band of interest. Instead, we would have to mix different sensors' outputs in a frequency-dependent way in order to reconstruct the true spot motion. In fact, the reconstruction of $x_{\rm spot}(t)$ is the most challenging component in mitigating the angular noise in the real aLIGO system currently. 

In our simulation, we assume the spot motion can be reconstructed by two sets of sensors. One set senses the spot motion via modulation-demodulation technique. Following LIGO convention, we will refer to these sensors as ``ADS'' sensors (with ADS standing for the alignment dithering system; \citealt{Buikema:20}). Specifically, we intentionally excite each mirror in angle at a known frequency and then demodulate the length readout (i.e., DARM output) at the same frequency. From Eq.~(\ref{eq:bilin_single_mirror}) we see that the demodulated signal is directly proportional to the spot motion on the test mass. However, this technique has a limited signal-to-noise ratio (SNR) because only a weak excitation is allowed in order to avoid saturation of actuators and sensors as well as up-conversion of the low-frequncy noise into the sensitivity ($>10\,{\rm Hz}$) band. Consequently, the ADS sensors are sensitive to only the very low frequency ($<0.1\,{\rm Hz}$) portion of the spot motion. 

The other set of sensors are known as ``optical levers'' and we will refer to them as OPLEVs following the LIGO convention~\citep{Black:10}. An OPLEV probes a mirror's alignment locally, which then allows us to infer the spot motion using Eq.~(\ref{eq:a2s}). However, the reference point drifts slowly and thus an OPLEV can only probe spot motion in the $\gtrsim 0.1 {\rm Hz}$ band. 
Consequently, we would need to combine both ADS and OPLEV sensors together to reconstruct the spot motion, as shown in the right panel of Fig.~\ref{fig:bilin_spot} where the coherence between each sensor and the simulated true spot motion is shown. We use the color orange and blue to respectively represent the coherence with an ADS sensor and a pair of OPLEVs (as the spot motion on a mirror depends on the angular motion of both mirrors forming the cavity). The multi-input-single-output (MISO) coherence using all the sensors is shown in the grey trace. In the legend, we also quote the root-mean-square (RMS) value of spot motion that is coherent with the a specific set of sensors (the true spot motion has an RMS of 1\,mm here). 
Note that the coherence is related to SNR at each frequency bin as 
\begin{equation}
    {\rm SNR}^2 = \frac{{\rm Coherence}}{1-{\rm Coherence}}.
    \label{eq:snr_vs_coh}
\end{equation}
This means that we would need to include at least 16 sensors (8 ADSs + 8 OPLEVs) in order to sense spot motions on the 4 test masses in both pitch and yaw over the entire band of interest. Together with 4 fast channels corresponding to the 4 hard mode feedback (as there are 2 ARM cavities and each cavity has 2 angular degrees of freedom), we include 20 auxiliary channels in total in our simulation of the angular noise.\footnote{The channels we simulate are in fact based on real LIGO auxiliary channels. The fast angular motion $\theta^{\rm (h)}(t)$ are contained in channels like \texttt{L1:ASC-CHARD\_P\_IN1\_DQ}. The slow channels are designed to mimic auxiliary channels like \texttt{L1:ASC-ADS\_PIT4\_DOF\_OUT\_DQ} for the ADS sensors, and \texttt{L1:SUS-ITMX\_L3\_OPLEV\_PIT\_OUT\_DQ} for the OPLEVs.} 

We note that in reality, more channels may be needed for the spot motion. In the right panel of Fig.~\ref{fig:bilin_spot}, while the simulated OPLEVs have comparable sensitivity to the real ones, we are nonetheless being optimistic about the SNR of the ADS sensors. As we will see later in Sec.~\ref{sec:snr_req_ads}, without sufficient SNR in the $<0.1\,{\rm Hz}$ band the subtraction performance will be significantly limited. Therefore, in order to achieve successful noise regression in reality, we would need either more low-frequency channels with a more complicated CNN structure, or more accurate sensing schemes for the $\lesssim 0.1\,{\rm Hz}$  spot motion. 


\section{General vs. specific CNN structures}
\label{sec:NN_structure}

\begin{figure}[h!]
\begin{center}
\includegraphics[width=0.8\textwidth]{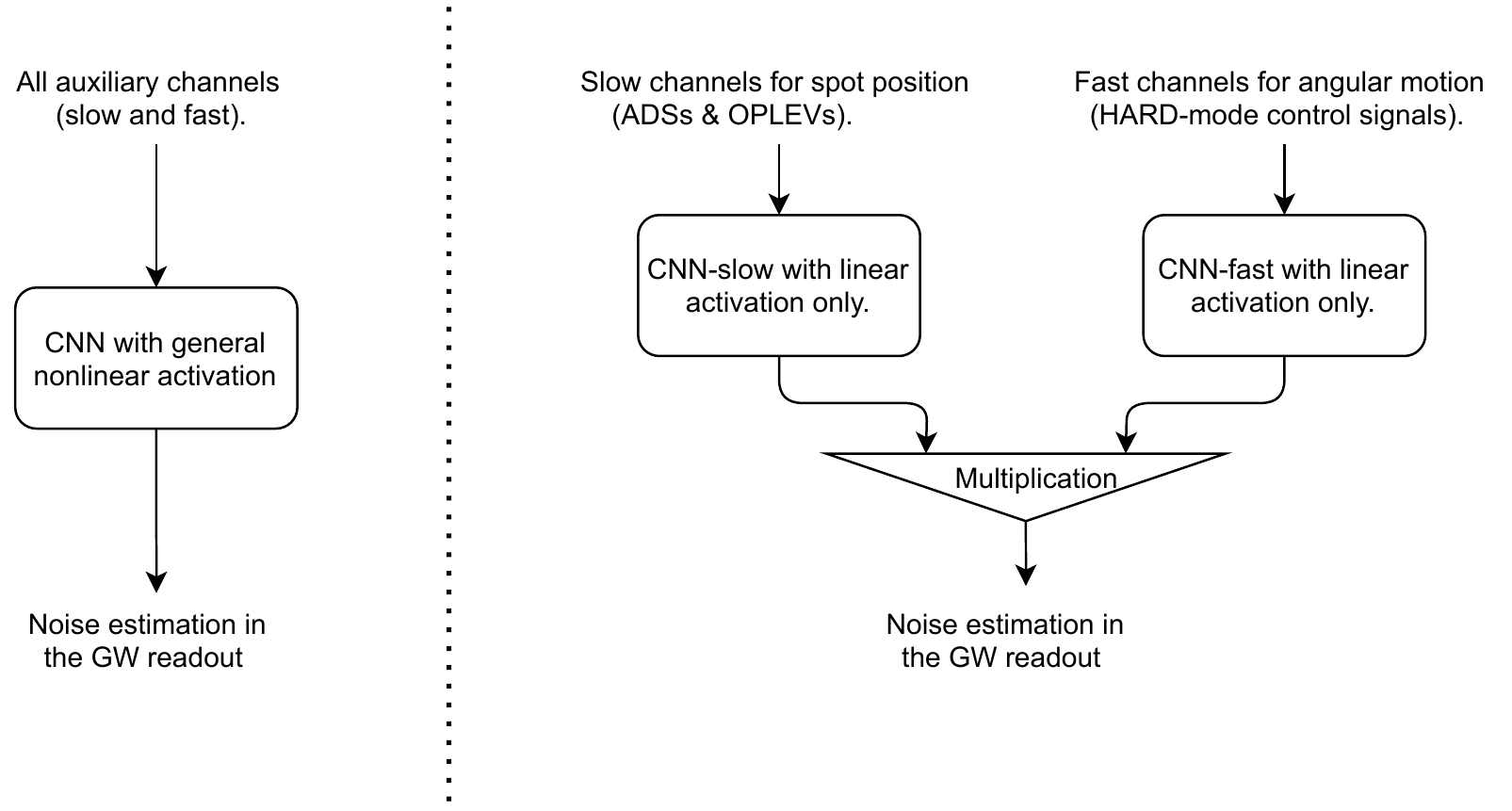}
\end{center}
\caption{Flow charts of a noise regression CNN utilizing a general structural (left; the parameters are detailed in Table~\ref{tab:CNN_reg}) and one that adopts an explicit ``slow$\times$fast'' structure (right; the parameters are summarized in Table~\ref{tab:CNN_sxf}). }\label{fig:flow_chart}
\end{figure}

\begin{figure}[h!]
\begin{center}
\includegraphics[width=\textwidth]{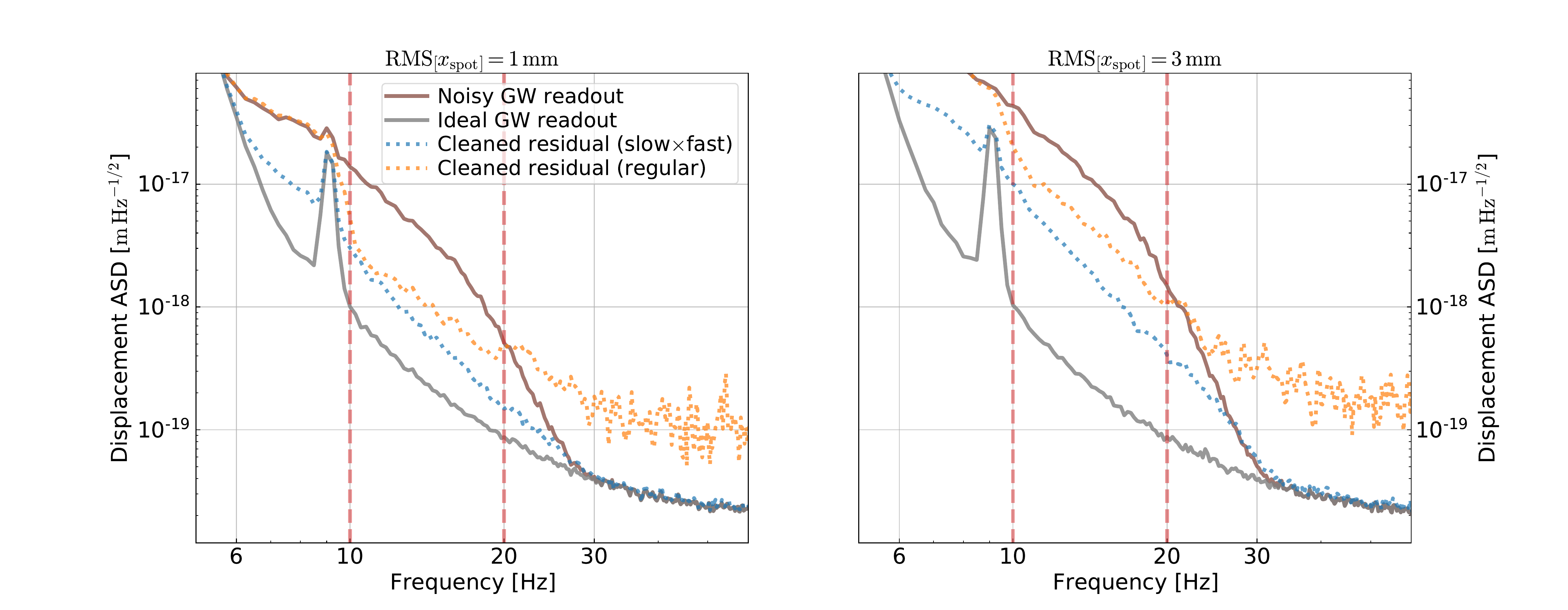}
\end{center}
\caption{Comparison of the noise subtraction results using the ``slow$\times$fast'' structure vs the general structure with ELU activation. In the left the spot motion has an RMS value of $1\,{\rm mm}$, which is the same as the training data set. The two vertical lines indicate the band used for computing the loss (i.e. the training band). Using the ``slow$\times$fast'' structure, we can  not only achieve a better subtraction in the training band but also avoid injecting extra contamination outside the band. In the right panel, we extrapolate the results on testing data with spot motion 3 times higher than the one used for training. The performance of the ``slow$\times$fast'' degrades less compared to the general one. }\label{fig:extrap_comparison}
\end{figure}

Having described our noise simulation, we now discuss the mitigation of the noise using ML-based CNN. Note that the noise regression problem is a problem of ``supervised learning'' and it can be tackled using ML by inputting time series of auxiliary channels into a CNN and training it to reconstruct the time series of the main GW readout (i.e., ``DARM''). Effectively, the training process is to help the CNN learn the noise coupling mechanism, Eq.~(\ref{eq:bilin_cav}), with appropriate, frequency-dependent combinations of different auxiliary channels so that when the training completes, the CNN can use the auxiliary channels to predict the noise in the GW readout. With the goal defined, the question then becomes the following. How do we design the structure of the CNN so that the regression performance is optimized?

One option is to use a general deep-filtering structure as illustrated in the left part of Fig.~\ref{fig:flow_chart}. Such a CNN should contain sufficiently many convolutional layers together with densely connected layers, and at least some of the layers should involve a nonlinear activation function such as ReLU or ELU at each layer's output. The convolutional layers would then behave similarly to finite-impulse-response filters, enabling frequency-dependent blending of different auxiliary sensors' time series (i.e., ``input'' to the CNN). Moreover, with sufficiently many layers with nonlinear activations, the nonlinearity involved in the coupling mechanism can be represented in a series-expansion sense. Such a deep-filtering CNN has the advantages of being straightforward and general. In fact, it would work without requiring any prior knowledge of the coupling mechanism (as long as the input contains all the information) and thus is particularly suitable to handle noise source with unknown couplings. 

On the other hand, for the bilinearly coupled angular noise described in Sec.~\ref{sec:sim_noise}, we nonetheless know how the noise propagates to the GW readout. The challenging part of its mitigation is the reconstruction of the spot motion on each test masses. In this scenario, we can in fact utilize our knowledge on the coupling mechanism and design a more specific CNN structure to tackle this problem as shown in the right part of Fig.~\ref{fig:flow_chart}. In particular, we divide the auxiliary channels into a slow set [ADSs and OPLEVs for the spot motion $x_{\rm spot}(t)$] and a fast set [HARD mode control signals for the angular motion $\theta^{(h)}(t)$]. Each set goes through a CNN (which will be referred to as ``CNN-slow'' and ``CNN-fast'', respectively) which requires only linear activation, as the nonlinearity involved in the problem is made explicit via the multiplication layer. For the rest of the Section, we will then demonstrate how the specific structure may improve the mitigation performance relative to the general one. As in aLIGO, the coupling mechanisms behind many noise sources are in fact established (see, e.g.,\citealt{Martynov:16} and \citealt{Buikema:20} and references therein; see also Sec.~\ref{sec:summary}), this Section further serves as a demonstration of the benefits of incorporating our knowledge on the instrument when considering noise regression problems.

To compare the performance of the two structures, we consider CNNs with hyper-parameters listed in Tables~\ref{tab:CNN_reg} and \ref{tab:CNN_sxf} respectively for the general and the ``slow$\times$fast'' structures. Note here the input is a 3-dimension array. The first dimension corresponding to the batch dimension (i.e., different realization of the simulated data). The second dimension corresponds to the number of input channels. It starts as 20 for the general structure, 16 for CNN-slow, 4 for CNN-fast, and its shape changes according to the ``output dimension'' column in Tables~\ref{tab:CNN_reg} and \ref{tab:CNN_sxf} in subsequent layers. Note that for the final layer it should be 1 corresponding to the main GW readout (i.e., target of the training). Lastly, the final dimension is corresponds to the temporal dimension and it is the axis along which the convolution is performed.  During the tuning of hyper-parameters, we intentionally keep the CNN using the general structure (Table~\ref{tab:CNN_reg}) to be similar to CNN-slow so that the results can be compared fairly (note both CNN-fast and the network after the multiplication require only simple structures with a small fraction of total trainable parameters). 

We utilize a custom loss function $\mathcal{L}$ based on the band-limited PSD of the GW readout as 
\begin{equation}
    \mathcal{L} = k\int_{f_1}^{f_2}  \frac{S_n(f)}{S_n^{(0)}(f)} \left(\frac{f}{f_1}\right)^\alpha df,
    \label{eq:loss}
\end{equation}
where $S_n$ is the PSD of the GW readout after the subtraction (i.e., the target channel) it is normalized by $S_n^{(0)}$, the original PSD before noise subtraction (which is fixed during the training). We have additionally include an power-law weighting with an index $\alpha$; we empirically set $\alpha=-1$ in our case. The loss is computed over the frequency band $[f_1, f_2]$ and we tune the overall gain $k$ such that the initial loss is about order unity. Note that Eq.~(\ref{eq:loss}) does not constrain the DC value of the GW readout. While such a DC offset in the GW readout does not directly affect the sensitivity to astrophysical sources, we nonetheless choose to avoid introducing any offsets during the noise regression. This is achieved by adding the regular mean-square error together with Eq.~(\ref{eq:loss}) to form our final loss function used during the training process. The relative weights is tuned such that the mean-square error contributes about $10\%$ to the total loss initially. 

Because the RMS value of, e.g., the spot motion contains physical information, we do not normalize each channel by its variance [which may be different for different datasets as the RMS of $x_{\rm spot}(t)$ varies over time]. Instead, we want each channel to be always normalized by the same constant. For this purpose, we calibrate the ADS channels to be in [mm], OPLEVs in [mrad], and the main GW readout (``DARM'') in [fm]. For the HARD-mode error signals, we calibrate them to [pm] and further multiply it by the open-loop gain to precondition it to be proportional to the physical $\theta^{(h)}(t)$.

For each CNN, we use 1536\,s of data sampled at 128 Hz for training (which we further divide into 8 batches during the training), 256\,s of data for validation, and finally 256\,s of data for testing the results. We train each CNN until the loss on the validation data plateaus. 

The subtraction results on the testing data are shown in the left panel of Fig.~\ref{fig:extrap_comparison}. Here the brown trace is the ASD including both the fundamental noise and the bilinear noise. The ASD of the fundamental noise is also shown separately in the grey trace; in the ideal case, a CNN should use the information stored in the auxiliary channels to reduce the brown ASD to the grey one. The ASDs of residual time series cleaned with our CNNs are shown in the dotted traces. Here we use the color olive and blue to respectively represent the results obtained from the general (Table~\ref{tab:CNN_reg}) and the ``slow$\times$fast'' (Table~\ref{tab:CNN_sxf}) CNNs. 

One question we want to address specifically is how well each CNN extrapolates. Particularly, one quantity we extrapolate is the frequency. While during the training we would like to focus on the frequency band where the noise contamination is the highest to get the best mitigation there, we also want to avoid injecting excess noise outside this band during the subtraction process. To test this point, we thus set the training band [i.e., the band where we compute the loss function in Eq.~(\ref{eq:loss})] to be $[f_1, f_2]{=}[10,\ 20]\,{\rm Hz}$, corresponding to the two red-vertical lines in Fig.~\ref{fig:extrap_comparison}.  As shown in the figure, inside the $[10,\ 20]\,{\rm Hz}$ band both the general and the ``slow$\times$fast'' CNNs have decent and comparable subtraction performance (about a factor of 10 at 10\,Hz; the ``slow$\times$fast'' CNN has a slightly better performance at other frequencies). Outside the training band, on the other hand, the ``slow$\times$fast'' CNN significantly outperforms the general one. Whereas the general CNN starts to inject excess noise into the GW readout almost immediately outside the training band, the one adopting the ``slow$\times$fast'' structure continues to reduce the noise for the entire band from 6\,Hz to 30\,Hz where the angular noise is above the fundamental limit. Above 30\,Hz, the ``slow$\times$fast'' CNN also avoids making the noise worse than the fundamental limit. We thus see that by utilizing our knowledge on the physical systems, we can achieve good extrapolation properties with respect to frequency. 

Meanwhile, we would also like to assess how the CNNs perform at different values of RMS of the spot motion. This is because in the real aLIGO system the spot motion is induced by the $<1\,{\rm Hz}$ seismic motion, which varies over time, and the CNN's performance would thus need to be robust against the variation in the spot motion. We address this point by simulating an additional 256 seconds of data with an RMS spot motion of 3\,mm, and then testing our CNNs on this dataset with 3 higher spot motion than the training dataset. The result is summarized in the right panel in Fig.~\ref{fig:extrap_comparison}. While the fractional noise reduction at 10\,Hz degrades for both CNNs, the degradation is less if one uses the ``slow$\times$fast'' structure than the general structure. 

\section{SNR requirements}
\label{sec:snr_req}
In the section above, we have considered a simple case where the total GW readout contains only a fundamental component and all the excess contamination is due to the angular noise. In reality, there are many other types of excess noises in the low-frequency part of the aLIGO sensitivity band whose information are not captured in the auxiliary channels we input to the CNN. Effectively, the presence of other contamination reduces the angular noise's SNR in the GW readout (i.e., the target channel; note here we treat the angular noise as the signal because it is what we want to remove by the CNN). We thus explore how the CNN's performance varies with respect to the GW readout in Sec.~\ref{sec:snr_req_darm}. 

Another thing we would like to explore is the SNR of ADS sensors for the very low frequency ($\lesssim 0.1\,{\rm Hz}$) spot motion. The coherence level shown in the right panel of Fig.~\ref{fig:bilin_spot} is likely to be an optimistic estimation for the $\lesssim 0.1\,{\rm Hz}$ band because sensing the spot motion in this band intrinsically challenging. Not only does the ADS sensors have high sensing noise as described in Sec.~\ref{sec:sim_noise}, but also the spot estimation could be biased by other contamination mechanisms.\footnote{For example, the power circulating in the arm cavity may be modulated by $\theta^{\rm (mir)}(t)$ due to both the residual angular motion at $<1\,{\rm Hz}$ and/or imperfections in the mirror coating. The power fluctuation could further lead to a longitudinal perturbation mitigated by radiation pressure in addition to the geometrical effect Eq.~(\ref{eq:bilin_single_mirror}). Thus the effective spot position inferred at a given dithering frequency could be an biased estimation for another frequency.} Therefore, we also assess the CNN's performance with respect to the SNR in the ADS sensors in Sec.~\ref{sec:snr_req_ads}. 

\subsection{In the GW readout}
\label{sec:snr_req_darm}

\begin{figure}[htb]
\begin{center}
\includegraphics[width=\textwidth]{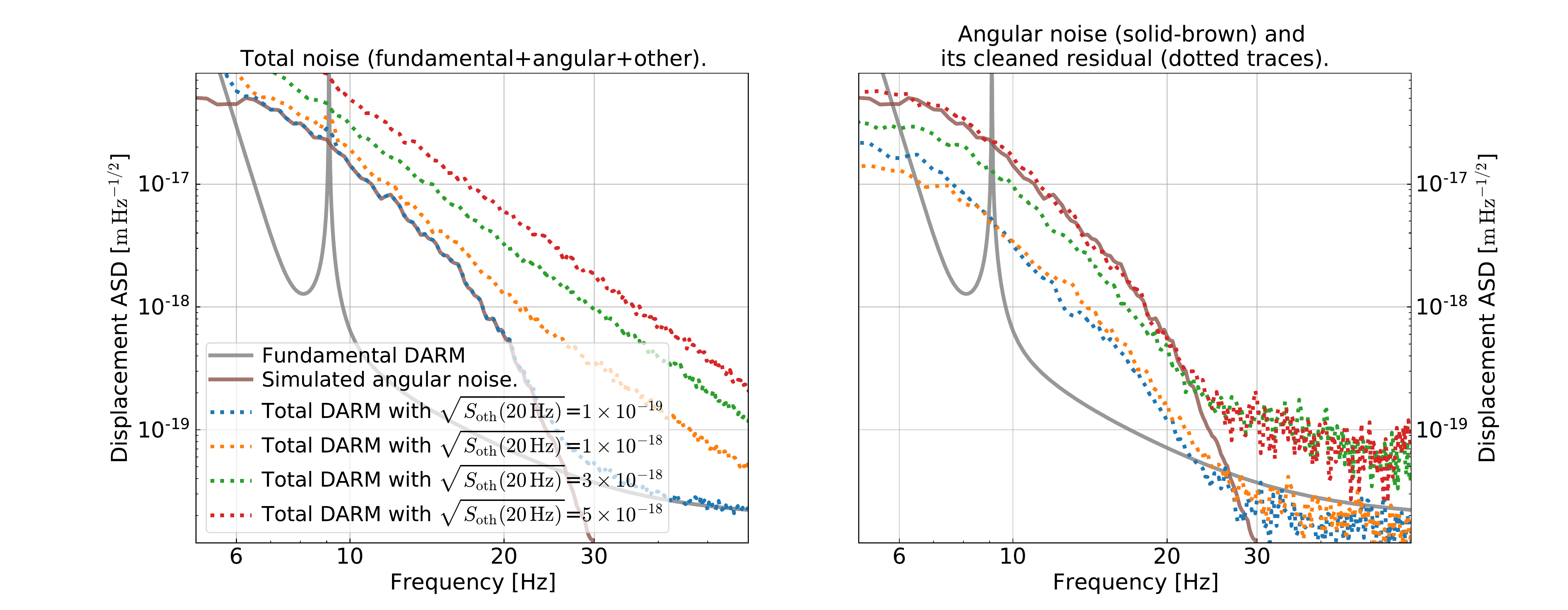}
\end{center}
\caption{Examining subtraction performance at different levels of SNR in the GW readout (i.e., ``DARM''). Here ``SNR'' means the ratio of the angular noise to the other, non-angular noises in DARM (including the fundamental noise and other technical noises whose ASD is denoted by $\sqrt{S_{\rm oth}}$). In the left panel, we show the DARM spectra at different values of other technical noises (dotted lines), while holding the angular noise fixed at the brown-solid trace. In the right panel, we show the subtraction residual of the angular noise (the lower the better). The color of each dotted trace corresponds to the same value of $\sqrt{S_{\rm oth}}$ as in the right panel. We see that the subtraction has similar performance when when the ${\rm SNR}\gtrsim 1$ (blue and orange traces), while it degrades quickly when ${\rm SNR}<1$ (green and red traces). }\label{fig:resi_diff_snr_in_darm}
\end{figure}

To modify the SNR of the simulated angular noise in the main GW readout, we simulate the total length fluctuation $\delta x_{\rm tot}$ in the GW readout as
\begin{equation}
    \delta x_{\rm tot}(t) = \delta x_{\rm fun}(t) + \delta x_{\rm ang}(t) + \delta x_{\rm oth}(t),
    \label{eq:dx_w_oth}
\end{equation}
where $\delta x_{\rm fun}$ corresponds the fundamental noise component (grey trace in Fig.~\ref{fig:bilin_darm_asd}), $\delta x_{\rm ang}$ is the angular noise we want to subtract remove (red trace in Fig.~\ref{fig:bilin_darm_asd}), and $\delta x_{\rm oth}$ are other types of contamination to DARM whose information is not contained in the auxiliary channels we input. For the other contamination, we further assume that it has an ASD given by $\sqrt{S_{\rm oth}(f)}\propto f^{-3}$. This way, we can control the SNR[=$\sqrt{S_{\rm ang}(f)}/\sqrt{S_{\rm fun}(f) + S_{\rm oth}(f)}$] of the bilinear noise 
by changing the overall magnitude of the other noise's ASD.

In the left panel of Fig.~\ref{fig:resi_diff_snr_in_darm}, we show the ASDs of the total DARM displacement, now including the contribution from  other contamination simulated according to Eq.~(\ref{eq:dx_w_oth}). We use dotted lines with different colors to indicate different levels of the other noise while the angular noise is held fixed as shown in the solid-brown trace in this section. For each value of $\sqrt{S_{\rm oth}(20\,{\rm Hz})}$, we regenerate 2048 seconds of data (1536 s for training, 256 s for validation, and 256 second for testing). 
We then train our CNN on the new data and check how much it could mitigate the bilinearly coupled angular noise (solid-brown trace in the left panel of Fig.~\ref{fig:resi_diff_snr_in_darm}). Here we focus on the CNN with the ``slow$\times$fast'' structure (Table~\ref{tab:CNN_sxf}) as it has a better performance than the general one (Sec.~\ref{sec:NN_structure}). The auxiliary channels are still assumed to have good sensitivity as shown in the right panel of Fig.~(\ref{fig:bilin_spot}). The training strategy follows the one outlined in Sec.~\ref{sec:NN_structure} except for that we now evaluate the loss, Eq.~(\ref{eq:loss}), over a broader band of $[f_1, f_2]=[8, 40]\,{\rm Hz}$. 

In the right panel, we show the ASDs of the residual time series after noise mitigation by our CNN. For presentation purpose, we show specifically the residual angular noise component in each dotted line. This is obtained by subtracting from the residual [i.e., the difference between the original $\delta x_{\rm tot} (t)$ and the one predicted by our CNN] further the $\delta x_{\rm fun}(t)$ and $\delta x_{\rm oth}(t)$ components when generating the plot, which is possible as we are dealing with the simulated data. Different colors correspond to different levels $\sqrt{S_{\rm oth}(20\,{\rm Hz})}$ and the correspondence is the same as in the left panel. 

As shown in the plot, when the angular noise has an ${\rm SNR}\gtrsim 1$ at each frequency bin in the GW readout, the CNN can achieve decent and consistent subtraction performance as shown by the blue and orange traces. At 10\,Hz, the angular noise could be reduced by about an order of magnitude. However, when the SNR drops below unity as shown in the green trace (with an SNR of about 0.5 at 10 Hz), the CNN could only marginally reduce the angular noise in the 10-20 Hz band, and it starts to inject excess contamination at higher frequencies. 

Therefore, for the CNN to be able to recognize the correct noise coupling mechanism, it would typically need an SNR of unity in the target channel. Nonetheless, we note this is not a \emph{necessary} condition. As we will discuss later in Sec.~\ref{sec:CL}, we may still be able to remove contaiminations in DARM with a quiet-time SNR sub-unity if we utilize active injection and curriculum learning.


\subsection{In the witness sensors}
\label{sec:snr_req_ads}

\begin{figure}[htb]
\begin{center}
\includegraphics[width=\textwidth]{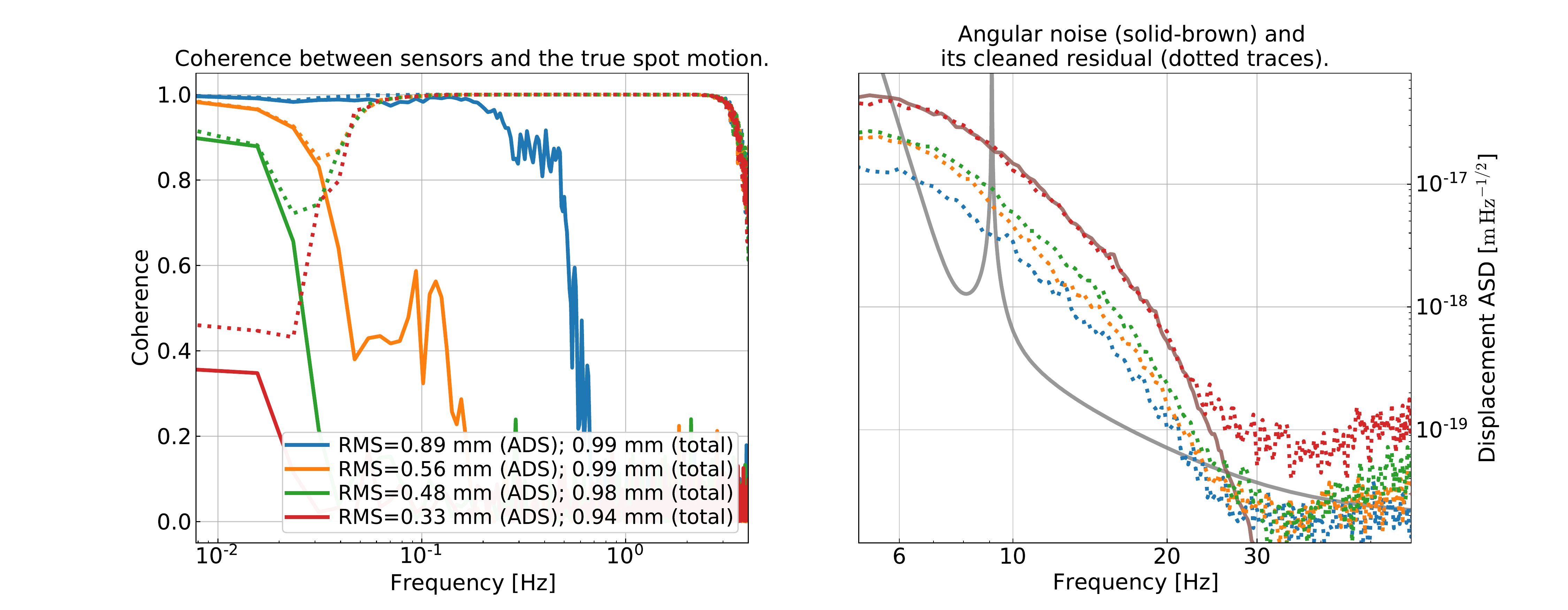}
\end{center}
\caption{Examining subtraction performance at different levels of SNR in the low-frequency ADS channels (which probes the true beam spot motion $\lesssim0.1\,{\rm Hz}$). In the left, we show the coherence between the ADS channels (solid traces) and the true spot motion on one of the test masses in pitch. As a reference, the total MISO coherence using two OPLEVs and one ADS channels as the witnesses is shown in the dashed traces. In the legend, we show the RMS of the spot motion coherent with the witnesses (either a single ADS or further combining it with two OPLEV sensors; the total spot RMS is 1.0\,mm in all cases). In the right, we show the subtraction residual of the angular noise (the lower the better) with the color of each dotted trace corresponds to the same level of sensor SNR as in the left. While we can reconstruct $\gtrsim 95\%$ in power of the true spot motion even in the red traces (the worst case considered), the subtraction degrades significantly as the SNR in the ADS channels reduces. It thus indicates the necessity of having good sensors covering the entire frequency band of interests (even the band contributes only a small fraction of the total RMS). }\label{fig:resi_diff_snr_in_ads}
\end{figure}

So far we have assumed the ADS sensors have good SNR of the true spot motion below 0.1\,Hz as indicated in the left panel of Fig.~\ref{fig:bilin_spot}. This may be an optimistic assumption in reality as discussed at the beginng of Sec.~\ref{sec:snr_req}. We thus explore here how does the SNR in the ADS channels (i.e., inputs or witnesses) affect the performance of the CNN. 

For this purpose, we vary the sensing noise in each ADS channel to modify its coherence with the true spot motion. This is indicate in the left panel of Fig.~\ref{fig:resi_diff_snr_in_ads}. We use the solid trace to represent the coherence between an ADS's output and the true spot motion it senses, and the dashed trace the MISO coherence if we further include a pair of OPLEVs. In the legend, we quote the RMS value of the spot motion that is coherent with the sensors (the true spot motion has an RMS of $1\,{\rm mm}$). Note the coherence is related to the SNR via Eq.~(\ref{eq:snr_vs_coh}). 

We repeat the training process now on data with noisy ADS sensors and the subtraction result is summarized in the right panel of Fig.~\ref{fig:resi_diff_snr_in_ads}. The color coding of each curve has the same meaning as in the left panel. We see that as the ADS sensors' sensitivity decreases, the noise subtraction performance degrades significantly. Take the green trace as an example. Even the MISO coherence is greater than at least 0.7 over the entire band of interest and the RMS of the spot motion coherent with the sensors (ADS + OPLEVs) almost matches the RMS of the true value, the amount of noise the CNN can subtract reduces by a factor of 2 compared to the blue trace. Our study thus suggests that the high-accuracy reconstruction of the beam spot on each test mass over the entire $<1\,{\rm Hz}$ band is crucial for the success of the mitigation of the angular noise. In fact, the spot reconstruction should a topic deserving dedicated studies on its own right.

\section{curriculum learning}
\label{sec:CL}

\begin{figure}[htb]
\begin{center}
\includegraphics[width=0.5\textwidth]{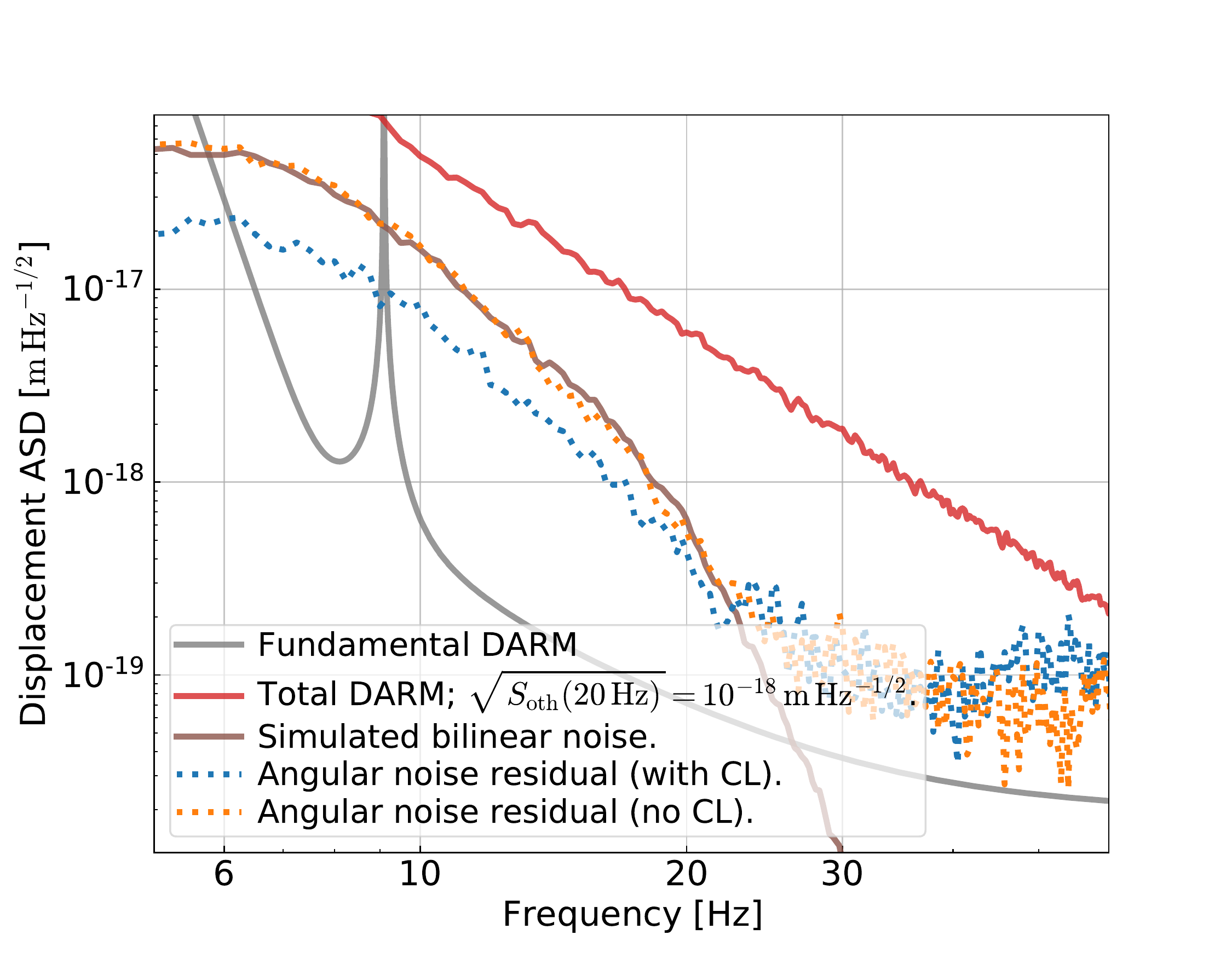}
\end{center}
\caption{Examining subtraction performance with or without curriculum learning (CL). Here we considered the same dataset as shown in the red traces in Fig.~\ref{fig:resi_diff_snr_in_darm}. I.e., while the angular noise (subtraction target) is high compared to the fundamental DARM noise, it is buried by other technical noises with an SNR<1. Without CL, no mitigation of the angular noise is achieved (as in Fig.~\ref{fig:resi_diff_snr_in_darm}). Using CL, on the other hand, allows the network to achieve better performance. This suggests that in reality, we can use active noise excitation and CL training to enhance the performance of noise subtraction on the quiet-time data.}\label{fig:cur_learn_snr_darm}
\end{figure}

As discussed in the previous section, in order for the CNN to be able to remove the angular noise, we would need to have good SNRs in both the target (the main GW readout) and the witness sensors (the ADS sensors). On the other hand, difference exists between the two scenarios. If the SNR is low in the witnesses (Sec.~\ref{sec:snr_req_ads}), it means we do not have enough information to recover the noise coupling and the only solution is to incorporate more sensors to recover the missing information. If the SNR is low only in the target (Sec.~\ref{sec:snr_req_darm}), then the fact that the CNN does not achieve a good subtraction is simply due to it not converging to the right physics during the training process. All the necessary information to predict the angular noise are still available. In this case, we show in this section that we can still have decent mitigation if we utilize curriculum learning (CL) techniques \citep{George:18a, George:18b}. 

The key idea behind CL is the following. We can first train the CNN first on datasets with high SNR in the target channel, which can be produced in reality with active injections. A high-SNR dataset is a simple task to be tackled and it helps the CNN to converge to the right physics initially. We can then gradually increase the task's difficulty by incorporating in data with lower excitation levels, and eventually the quiet-time data (i.e., data without any active injections). Since the CNN is guided through the training process, it stays close to the true physics and can thus predict the desired target even using quite-time data. This process is especially suitable to be combined with a physically-inspired CNN structure (such as the ``slow$\times$fast'' one) that has a good extrapolation property. 

We demonstrate this point here by first training our CNN on data corresponding to the blue traces in Fig.~\ref{fig:resi_diff_snr_in_darm}.\footnote{In reality, $\sqrt{S_{\rm oth}}$ is approximately fixed and to create high-SNR data we should instead inject excess motion to, e.g., $\theta^{(h)}$. Nonetheless, since what matters most is the SNR of the bilinear noise in the GW readout, here we simply reuse the existing data. } After its convergence, we then incorporate into the training set the data corresponding to the olive traces (as different ``batches'') and continue training the CNN obtained from the first step until its validation loss plateaus.  In the last step, we further include data corresponding to the red trace (with the lowest SNR in the GW readout) and retrain the CNN from previous step to convergence. 

After the CL training, we test the resultant CNN on the 256-s testing data for the red trace in Fig.~\ref{fig:resi_diff_snr_in_darm} and the result is presented in Fig.~\ref{fig:cur_learn_snr_darm} in the dotted-blue trace. As a comparison, we also shown the subtraction result without curriculum learning in the dotted-orange trace (using the CNN obtained in Sec.~\ref{sec:snr_req_darm}). Whereas the CNN could not mitigate any of the angular noise without CL, we see that with the help of CL the CNN can reduce the noise by a factor of 2 at around 10 Hz. It thus demonstrates that CL could be a useful training strategy to help the CNN remove noise that has a low SNR in the GW readout (yet it can still be high compared to the fundamental limit and thus needs to be tackled). 
%




\section{Summary and discussion}
\label{sec:summary}
In this work, we explored how we may use ML-based CNNs to improve the sub-60 Hz sensitivity of aLIGO. Here we focused specifically the bilinearly-coupled angular noise, which is one of the limiting noise sources in the 10-30 Hz band. Using simulated data with characteristics similar to the real aLIGO system (Sec.~\ref{sec:sim_noise}), we explored various factors affecting a CNN's performance. One of the most important factors is to utilize our knowledge into the design of CNN structures. Specifically, a ``slow$\times$fast'' structure is particularly suitable to mitigate the angular noise as it incorporates the nonlinearity involved in the problem in an explicit way and thus leads to a good extrapolation properties (Sec.~\ref{sec:NN_structure}). We further explored the SNR requirements for the CNN to converge in both the target (GW readout; Sec.~\ref{sec:snr_req_darm}) and in the witness sensors (ADS sensors; Sec.~\ref{sec:snr_req_ads}). To overcome the lack of SNR in the witnesses, it would require improving the sensing technology and/or including more sensors to recover the information. On the other hand, when the SNR is only low in the target, we demonstrated in Sec.~\ref{sec:CL} that CL training can be used to guide the CNN to converge to the right place.

While in this work we focused on the angular noise that couples to DARM according to Eq.~(\ref{eq:bilin_single_mirror}), we note that the coupling mechanisms of many other noises in LIGO share the same bilinear structure and can be modeled as a fast channel modulated by a slow one (i.e., the product of the two). For example, the signal-recycling cavity's length fluctuation (a fast channel) couples to DARM with a slow modulation due to low-frequency variations in the DARM offset. Other noise like light scattering, when expanded into a Taylor series, will also have terms similar to Eq.~(\ref{eq:bilin_single_mirror}) serving as the lowest-order nonlinear terms. Therefore, the ``slow$\times$fast'' CNN structure we present in Sec.~\ref{sec:NN_structure} will have broad applications in LIGO noise mitigation beyond just the angular noise. We plan to explore this point more in future studies. 

We note that the CL result shown in Fig.~\ref{fig:cur_learn_snr_darm}, while having an improved performance compared to the case without CL, has not yet reached a level comparable to the case where it converges to the right physics (e.g., the blue and orange traces in the right panel of Fig.~\ref{fig:resi_diff_snr_in_darm}). It thus indicates that the detailed CL steps has rooms for further optimization, which we defer to be explored by future studies. 

Throughout the analysis, we have focused on using simulated data. As we have mentioned in the main text, mitigating the angular noise in the real aLIGO system is more challenging. This is because there are many other noise sources  at the same frequency band (including ones that are not yet captured in the current noise budgeting; \citealt{Martynov:16, Buikema:20}), limiting the SNR of the angular noise in the GW readout. Moreover, to reconstruct $x_{\rm spot}(t)$ it is likely requiring more sensors than just the ADSs and OPLEVs. We also ignored potential transient (``glitches'') in both the GW readout and witness sensors, which is another crucial application of ML in GW astrophysics on its own right (see, e.g., \citealt{Cuoco:20} and references therein). On the real data, our CNN has not yet achieved significant broadband reduction of nonlinearly coupled noises, yet it shows promising signs such as removing the sidebands on the peaks to create the ADS signals~\citep{Yu:21z}. We plan to investigate this further in future studies, especially combining it with an optimized CL training strategy. 
Meanwhile, LIGO has released a 3-hour data stretch in its second observing run including major auxiliary channels.\footnote{The data can be downloaded at \url{https://www.gw-openscience.org/auxiliary/GW170814/}.} We would like to thus encourage interested readers to use either the CNN structures we proposed in this work or original CNN structures to help the further improvements of aLIGO's sensitivity.

\section*{Conflict of Interest Statement}

The authors declare that the research was conducted in the absence of any commercial or financial relationships that could be construed as a potential conflict of interest.

\section*{Author Contributions}

HY performed the training and testing of the CNNs. RXA supervised the research. 

\section*{Funding}
HY is supported by the Sherman Fairchild Foundation. RXA is supported by NSF Grants No. PHY-1764464. The authors gratefully acknowledge the computational resources provided by the LIGO Laboratory and supported by NSF Grants No. PHY-0757058 and No. PHY-0823459.

\section*{Acknowledgments}
We thank Gabriele Vajente and Szabolcs Marka for helpful discussions and comments during the preparation of this manuscript, and we acknowledge feedback from fellow participants in the LIGO Machine Learning Algorithms Call.


\section*{Data Availability Statement}
The simulated datasets for this study will be shared on reasonable request to the corresponding author.

\bibliographystyle{frontiersinHLTH_FPHY} 
\bibliography{ref}

\begin{thebibliography}{73}
\expandafter\ifx\csname natexlab\endcsname\relax\def\natexlab#1{#1}\fi
\expandafter\ifx\csname urlstyle\endcsname\relax
  \expandafter\ifx\csname doi\endcsname\relax
  \def\doi#1{doi:\discretionary{}{}{}#1}\fi \else
  \expandafter\ifx\csname doi\endcsname\relax
  \def\doi{doi:\discretionary{}{}{}\begingroup \urlstyle{rm}\Url}\fi \fi
\expandafter\ifx\csname selectlanguage\endcsname\relax
  \def\selectlanguage#1{}\fi

\bibitem[{Abbott et~al.(2016)}]{Abbott:2016blz}
Abbott BP, et~al.
\newblock {Observation of Gravitational Waves from a Binary Black Hole Merger}.
\newblock {\em Phys. Rev. Lett.\/} {\bf 116} (2016) 061102.
\newblock \doi{10.1103/PhysRevLett.116.061102}.

\bibitem[{{LIGO Scientific Collaboration} and et~al.(2015)}]{LSC:15}
{LIGO Scientific Collaboration}, et~al.
\newblock {Advanced LIGO}.
\newblock {\em Classical and Quantum Gravity\/} {\bf 32} (2015) 074001.
\newblock \doi{10.1088/0264-9381/32/7/074001}.

\bibitem[{Acernese et~al.(2015)}]{TheVirgo:2014hva}
Acernese F, et~al.
\newblock {Advanced Virgo: a second-generation interferometric gravitational
  wave detector}.
\newblock {\em Class. Quant. Grav.\/} {\bf 32} (2015) 024001.
\newblock \doi{10.1088/0264-9381/32/2/024001}.

\bibitem[{{Kagra Collaboration} and et~al.(2019)}]{Akutsu:2018axf}
{Kagra Collaboration}, et~al.
\newblock {KAGRA: 2.5 generation interferometric gravitational wave detector}.
\newblock {\em Nature Astronomy\/} {\bf 3} (2019) 35--40.
\newblock \doi{10.1038/s41550-018-0658-y}.

\bibitem[{{Abbott} et~al.(2019{\natexlab{a}}){Abbott}, {Abbott}, {Abbott},
  {Abraham}, {Acernese}, {Ackley} et~al.}]{LIGOScientific:2018mvr}
{Abbott} BP, {Abbott} R, {Abbott} TD, {Abraham} S, {Acernese} F, {Ackley} K,
  et~al.
\newblock {GWTC-1: A Gravitational-Wave Transient Catalog of Compact Binary
  Mergers Observed by LIGO and Virgo during the First and Second Observing
  Runs}.
\newblock {\em Physical Review X\/} {\bf 9} (2019{\natexlab{a}}) 031040.
\newblock \doi{10.1103/PhysRevX.9.031040}.

\bibitem[{{Abbott} et~al.(2020){Abbott}, {Abbott}, {Abraham}, {Acernese},
  {Ackley}, {Adams} et~al.}]{Abbott:2020niy}
{Abbott} R, {Abbott} TD, {Abraham} S, {Acernese} F, {Ackley} K, {Adams} A,
  et~al.
\newblock {GWTC-2: Compact Binary Coalescences Observed by LIGO and Virgo
  During the First Half of the Third Observing Run}.
\newblock {\em arXiv e-prints\/}  (2020) arXiv:2010.14527.

\bibitem[{{Tse} and et~al.(2019)}]{Tse:19}
{Tse} M, et~al.
\newblock {Quantum-Enhanced Advanced LIGO Detectors in the Era of
  Gravitational-Wave Astronomy}.
\newblock {\em \prl\/} {\bf 123} (2019) 231107.
\newblock \doi{10.1103/PhysRevLett.123.231107}.

\bibitem[{{Martynov} et~al.(2016){Martynov}, {Hall}, {Abbott}, {Abbott},
  {Abbott}, {Adams} et~al.}]{Martynov:16}
{Martynov} DV, {Hall} ED, {Abbott} BP, {Abbott} R, {Abbott} TD, {Adams} C,
  et~al.
\newblock {Sensitivity of the Advanced LIGO detectors at the beginning of
  gravitational wave astronomy}.
\newblock {\em \prd\/} {\bf 93} (2016) 112004.
\newblock \doi{10.1103/PhysRevD.93.112004}.

\bibitem[{{Buikema} et~al.(2020){Buikema}, {Cahillane}, {Mansell}, {Blair},
  {Abbott}, {Adams} et~al.}]{Buikema:20}
{Buikema} A, {Cahillane} C, {Mansell} GL, {Blair} CD, {Abbott} R, {Adams} C,
  et~al.
\newblock {Sensitivity and performance of the Advanced LIGO detectors in the
  third observing run}.
\newblock {\em \prd\/} {\bf 102} (2020) 062003.
\newblock \doi{10.1103/PhysRevD.102.062003}.

\bibitem[{{Cannon} et~al.(2012){Cannon}, {Cariou}, {Chapman},
  {Crispin-Ortuzar}, {Fotopoulos}, {Frei} et~al.}]{Cannon:12}
{Cannon} K, {Cariou} R, {Chapman} A, {Crispin-Ortuzar} M, {Fotopoulos} N,
  {Frei} M, et~al.
\newblock {Toward Early-warning Detection of Gravitational Waves from Compact
  Binary Coalescence}.
\newblock {\em \apj\/} {\bf 748} (2012) 136.
\newblock \doi{10.1088/0004-637X/748/2/136}.

\bibitem[{{Abbott} et~al.(2019{\natexlab{b}}){Abbott}, {Abbott}, {Abbott},
  {Abraham}, {Acernese}, {Ackley} et~al.}]{LSC:19}
{Abbott} BP, {Abbott} R, {Abbott} TD, {Abraham} S, {Acernese} F, {Ackley} K,
  et~al.
\newblock {Low-latency Gravitational-wave Alerts for Multimessenger Astronomy
  during the Second Advanced LIGO and Virgo Observing Run}.
\newblock {\em \apj\/} {\bf 875} (2019{\natexlab{b}}) 161.
\newblock \doi{10.3847/1538-4357/ab0e8f}.

\bibitem[{{Sachdev} et~al.(2020){Sachdev}, {Magee}, {Hanna}, {Cannon},
  {Singer}, {SK} et~al.}]{Sachdev:20}
{Sachdev} S, {Magee} R, {Hanna} C, {Cannon} K, {Singer} L, {SK} JR, et~al.
\newblock {An Early-warning System for Electromagnetic Follow-up of
  Gravitational-wave Events}.
\newblock {\em \apjl\/} {\bf 905} (2020) L25.
\newblock \doi{10.3847/2041-8213/abc753}.

\bibitem[{{Chu} et~al.(2020){Chu}, {Kovalam}, {Wen}, {Slaven-Blair}, {Bosveld},
  {Chen} et~al.}]{Chu:20}
{Chu} Q, {Kovalam} M, {Wen} L, {Slaven-Blair} T, {Bosveld} J, {Chen} Y, et~al.
\newblock {The SPIIR online coherent pipeline to search for gravitational waves
  from compact binary coalescences}.
\newblock {\em arXiv e-prints\/}  (2020) arXiv:2011.06787.

\bibitem[{{Yu} et~al.(2021){Yu}, {Adhikari}, {Magee}, {Sachdev}, and
  {Chen}}]{Yu:21c}
{Yu} H, {Adhikari} RX, {Magee} R, {Sachdev} S, {Chen} Y.
\newblock {Early warning of coalescing neutron-star and neutron-star-black-hole
  binaries from nonstationary noise background using neural networks}.
\newblock {\em arXiv e-prints\/}  (2021) arXiv:2104.09438.

\bibitem[{{Mandel} et~al.(2008){Mandel}, {Brown}, {Gair}, and
  {Miller}}]{Mandel:08}
{Mandel} I, {Brown} DA, {Gair} JR, {Miller} MC.
\newblock {Rates and Characteristics of Intermediate Mass Ratio Inspirals
  Detectable by Advanced LIGO}.
\newblock {\em \apj\/} {\bf 681} (2008) 1431--1447.
\newblock \doi{10.1086/588246}.

\bibitem[{{Graff} et~al.(2015){Graff}, {Buonanno}, and
  {Sathyaprakash}}]{Graff:15}
{Graff} PB, {Buonanno} A, {Sathyaprakash} BS.
\newblock {Missing Link: Bayesian detection and measurement of
  intermediate-mass black-hole binaries}.
\newblock {\em \prd\/} {\bf 92} (2015) 022002.
\newblock \doi{10.1103/PhysRevD.92.022002}.

\bibitem[{{Veitch} et~al.(2015){Veitch}, {P{\"u}rrer}, and
  {Mandel}}]{Veitch:15}
{Veitch} J, {P{\"u}rrer} M, {Mandel} I.
\newblock {Measuring Intermediate-Mass Black-Hole Binaries with Advanced
  Gravitational Wave Detectors}.
\newblock {\em \prl\/} {\bf 115} (2015) 141101.
\newblock \doi{10.1103/PhysRevLett.115.141101}.

\bibitem[{{LIGO Scientific Collaboration} and {Virgo
  Collaboration}(2020)}]{GW190521}
{LIGO Scientific Collaboration}, {Virgo Collaboration}.
\newblock {GW190521: A Binary Black Hole Merger with a Total Mass of 150
  M$_\odot$}.
\newblock {\em \prl\/} {\bf 125} (2020) 101102.
\newblock \doi{10.1103/PhysRevLett.125.101102}.

\bibitem[{{Romero-Shaw} et~al.(2019){Romero-Shaw}, {Lasky}, and
  {Thrane}}]{Romero-Shaw:19}
{Romero-Shaw} IM, {Lasky} PD, {Thrane} E.
\newblock {Searching for eccentricity: signatures of dynamical formation in the
  first gravitational-wave transient catalogue of LIGO and Virgo}.
\newblock {\em \mnras\/} {\bf 490} (2019) 5210--5216.
\newblock \doi{10.1093/mnras/stz2996}.

\bibitem[{Abbott et~al.(2019)}]{LSC:19b}
Abbott BP, et~al.
\newblock {Search for Eccentric Binary Black Hole Mergers with Advanced LIGO
  and Advanced Virgo during their First and Second Observing Runs}.
\newblock {\em Astrophys. J.\/} {\bf 883} (2019) 149.
\newblock \doi{10.3847/1538-4357/ab3c2d}.

\bibitem[{{Yu} et~al.(2018){Yu}, {Martynov}, {Vitale}, {Evans}, {Shoemaker},
  {Barr} et~al.}]{Yu:18}
{Yu} H, {Martynov} D, {Vitale} S, {Evans} M, {Shoemaker} D, {Barr} B, et~al.
\newblock {Prospects for Detecting Gravitational Waves at 5 Hz with
  Ground-Based Detectors}.
\newblock {\em \prl\/} {\bf 120} (2018) 141102.
\newblock \doi{10.1103/PhysRevLett.120.141102}.

\bibitem[{{Driggers} et~al.(2019){Driggers}, {Vitale}, {Lundgren}, {Evans},
  {Kawabe}, {Dwyer} et~al.}]{Driggers:19}
{Driggers} JC, {Vitale} S, {Lundgren} AP, {Evans} M, {Kawabe} K, {Dwyer} SE,
  et~al.
\newblock {Improving astrophysical parameter estimation via offline noise
  subtraction for Advanced LIGO}.
\newblock {\em \prd\/} {\bf 99} (2019) 042001.
\newblock \doi{10.1103/PhysRevD.99.042001}.

\bibitem[{{Davis} et~al.(2019){Davis}, {Massinger}, {Lundgren}, {Driggers},
  {Urban}, and {Nuttall}}]{Davis:19}
{Davis} D, {Massinger} T, {Lundgren} A, {Driggers} JC, {Urban} AL, {Nuttall} L.
\newblock {Improving the sensitivity of Advanced LIGO using noise subtraction}.
\newblock {\em Classical and Quantum Gravity\/} {\bf 36} (2019) 055011.
\newblock \doi{10.1088/1361-6382/ab01c5}.

\bibitem[{{Ormiston} et~al.(2020){Ormiston}, {Nguyen}, {Coughlin}, {Adhikari},
  and {Katsavounidis}}]{Ormiston:20}
{Ormiston} R, {Nguyen} T, {Coughlin} M, {Adhikari} RX, {Katsavounidis} E.
\newblock {Noise reduction in gravitational-wave data via deep learning}.
\newblock {\em Physical Review Research\/} {\bf 2} (2020) 033066.
\newblock \doi{10.1103/PhysRevResearch.2.033066}.

\bibitem[{{Vajente} et~al.(2020){Vajente}, {Huang}, {Isi}, {Driggers},
  {Kissel}, {Szczepa{\'n}czyk} et~al.}]{Vajente:20}
{Vajente} G, {Huang} Y, {Isi} M, {Driggers} JC, {Kissel} JS, {Szczepa{\'n}czyk}
  MJ, et~al.
\newblock {Machine-learning nonstationary noise out of gravitational-wave
  detectors}.
\newblock {\em \prd\/} {\bf 101} (2020) 042003.
\newblock \doi{10.1103/PhysRevD.101.042003}.

\bibitem[{{Mogushi} et~al.(2021){Mogushi}, {Quitzow-James}, {Cavagli{\`a}},
  {Kulkarni}, and {Hayes}}]{Mogushi:21}
{Mogushi} K, {Quitzow-James} R, {Cavagli{\`a}} M, {Kulkarni} S, {Hayes} F.
\newblock {NNETFIX: An artificial neural network-based denoising engine for
  gravitational-wave signals}.
\newblock {\em arXiv e-prints\/}  (2021) arXiv:2101.04712.

\bibitem[{{Baltus} et~al.(2021){Baltus}, {Janquart}, {Lopez}, {Reza},
  {Caudill}, and {Cudell}}]{Baltus:21}
{Baltus} G, {Janquart} J, {Lopez} M, {Reza} A, {Caudill} S, {Cudell} JR.
\newblock {Convolutional neural networks for the detection of the early
  inspiral of a gravitational-wave signal}.
\newblock {\em arXiv e-prints\/}  (2021) arXiv:2104.00594.

\bibitem[{{Huerta} et~al.(2020){Huerta}, {Khan}, {Huang}, {Tian}, {Levental},
  {Chard} et~al.}]{Huerta:20}
{Huerta} EA, {Khan} A, {Huang} X, {Tian} M, {Levental} M, {Chard} R, et~al.
\newblock {Confluence of Artificial Intelligence and High Performance Computing
  for Accelerated, Scalable and Reproducible Gravitational Wave Detection}.
\newblock {\em arXiv e-prints\/}  (2020) arXiv:2012.08545.

\bibitem[{{Krastev}(2020)}]{Krastev:20}
{Krastev} PG.
\newblock {Real-time detection of gravitational waves from binary neutron stars
  using artificial neural networks}.
\newblock {\em Physics Letters B\/} {\bf 803} (2020) 135330.
\newblock \doi{10.1016/j.physletb.2020.135330}.

\bibitem[{{Chan} et~al.(2020){Chan}, {Heng}, and {Messenger}}]{Chan:20}
{Chan} ML, {Heng} IS, {Messenger} C.
\newblock {Detection and classification of supernova gravitational wave
  signals: A deep learning approach}.
\newblock {\em \prd\/} {\bf 102} (2020) 043022.
\newblock \doi{10.1103/PhysRevD.102.043022}.

\bibitem[{{Dreissigacker} and {Prix}(2020)}]{Dreissigacker:20}
{Dreissigacker} C, {Prix} R.
\newblock {Deep-learning continuous gravitational waves: Multiple detectors and
  realistic noise}.
\newblock {\em \prd\/} {\bf 102} (2020) 022005.
\newblock \doi{10.1103/PhysRevD.102.022005}.

\bibitem[{{Wong} et~al.(2020){Wong}, {Ng}, and {Berti}}]{Wong:20}
{Wong} KWK, {Ng} KKY, {Berti} E.
\newblock {Gravitational-wave signal-to-noise interpolation via neural
  networks}.
\newblock {\em arXiv e-prints\/}  (2020) arXiv:2007.10350.

\bibitem[{{Sch{\"a}fer} et~al.(2020){Sch{\"a}fer}, {Ohme}, and
  {Nitz}}]{Schafer:20}
{Sch{\"a}fer} MB, {Ohme} F, {Nitz} AH.
\newblock {Detection of gravitational-wave signals from binary neutron star
  mergers using machine learning}.
\newblock {\em \prd\/} {\bf 102} (2020) 063015.
\newblock \doi{10.1103/PhysRevD.102.063015}.

\bibitem[{{Bayley} et~al.(2020){Bayley}, {Messenger}, and {Woan}}]{Bayley:20}
{Bayley} J, {Messenger} C, {Woan} G.
\newblock {Robust machine learning algorithm to search for continuous
  gravitational waves}.
\newblock {\em \prd\/} {\bf 102} (2020) 083024.
\newblock \doi{10.1103/PhysRevD.102.083024}.

\bibitem[{{Wei} and {Huerta}(2021)}]{Wei:21c}
{Wei} W, {Huerta} EA.
\newblock {Deep learning for gravitational wave forecasting of neutron star
  mergers}.
\newblock {\em Physics Letters B\/} {\bf 816} (2021) 136185.
\newblock \doi{10.1016/j.physletb.2021.136185}.

\bibitem[{{Chang} et~al.(2021){Chang}, {Onken}, {Wolf}, {Luvaul}, {M{\"o}ller},
  {Scalzo} et~al.}]{Chang:21}
{Chang} SW, {Onken} CA, {Wolf} C, {Luvaul} L, {M{\"o}ller} A, {Scalzo} R,
  et~al.
\newblock {SkyMapper optical follow-up of gravitational wave triggers: Alert
  science data pipeline and LIGO/Virgo O3 run}.
\newblock {\em \pasa\/} {\bf 38} (2021) e024.
\newblock \doi{10.1017/pasa.2021.17}.

\bibitem[{{Yan} et~al.(2021){Yan}, {Avagyan}, {Colgan}, {Veske}, {Bartos},
  {Wright} et~al.}]{Yan:21}
{Yan} J, {Avagyan} M, {Colgan} RE, {Veske} D, {Bartos} I, {Wright} J, et~al.
\newblock {Generalized Approach to Matched Filtering using Neural Networks}.
\newblock {\em arXiv e-prints\/}  (2021) arXiv:2104.03961.

\bibitem[{{Saiz-P{\'e}rez} et~al.(2021){Saiz-P{\'e}rez}, {Torres-Forn{\'e}},
  and {Font}}]{SaizPerez:21}
{Saiz-P{\'e}rez} A, {Torres-Forn{\'e}} A, {Font} JA.
\newblock {Classification of the core-collapse supernova explosion mechanism
  with learned dictionaries}.
\newblock {\em arXiv e-prints\/}  (2021) arXiv:2110.12941.

\bibitem[{{Chatterjee} et~al.(2021){Chatterjee}, {Wen}, {Diakogiannis}, and
  {Vinsen}}]{Chatterjee:21}
{Chatterjee} C, {Wen} L, {Diakogiannis} F, {Vinsen} K.
\newblock {Extraction of binary black hole gravitational wave signals from
  detector data using deep learning}.
\newblock {\em \prd\/} {\bf 104} (2021) 064046.
\newblock \doi{10.1103/PhysRevD.104.064046}.

\bibitem[{{Mishra} et~al.(2021){Mishra}, {O'Brien}, {Gayathri},
  {Szczepa{\'n}czyk}, {Bhaumik}, {Bartos} et~al.}]{Mishra:21}
{Mishra} T, {O'Brien} B, {Gayathri} V, {Szczepa{\'n}czyk} M, {Bhaumik} S,
  {Bartos} I, et~al.
\newblock {Optimization of model independent gravitational wave search for
  binary black hole mergers using machine learning}.
\newblock {\em \prd\/} {\bf 104} (2021) 023014.
\newblock \doi{10.1103/PhysRevD.104.023014}.

\bibitem[{{L{\'o}pez} et~al.(2021){L{\'o}pez}, {Di Palma}, {Drago},
  {Cerd{\'a}-Dur{\'a}n}, and {Ricci}}]{Lopez:21}
{L{\'o}pez} M, {Di Palma} I, {Drago} M, {Cerd{\'a}-Dur{\'a}n} P, {Ricci} F.
\newblock {Deep learning for core-collapse supernova detection}.
\newblock {\em \prd\/} {\bf 103} (2021) 063011.
\newblock \doi{10.1103/PhysRevD.103.063011}.

\bibitem[{{Beheshtipour} and {Papa}(2021)}]{Beheshtipour:21}
{Beheshtipour} B, {Papa} MA.
\newblock {Deep learning for clustering of continuous gravitational wave
  candidates. II. Identification of low-SNR candidates}.
\newblock {\em \prd\/} {\bf 103} (2021) 064027.
\newblock \doi{10.1103/PhysRevD.103.064027}.

\bibitem[{{Marianer} et~al.(2021){Marianer}, {Poznanski}, and
  {Prochaska}}]{Marianer:21}
{Marianer} T, {Poznanski} D, {Prochaska} JX.
\newblock {A semisupervised machine learning search for never-seen
  gravitational-wave sources}.
\newblock {\em \mnras\/} {\bf 500} (2021) 5408--5419.
\newblock \doi{10.1093/mnras/staa3550}.

\bibitem[{{Gabbard} et~al.(2019){Gabbard}, {Messenger}, {Heng}, {Tonolini}, and
  {Murray-Smith}}]{Gabbard:19}
{Gabbard} H, {Messenger} C, {Heng} IS, {Tonolini} F, {Murray-Smith} R.
\newblock {Bayesian parameter estimation using conditional variational
  autoencoders for gravitational-wave astronomy}.
\newblock {\em arXiv e-prints\/}  (2019) arXiv:1909.06296.

\bibitem[{{Chua} and {Vallisneri}(2020)}]{Chua:20}
{Chua} AJK, {Vallisneri} M.
\newblock {Learning Bayesian Posteriors with Neural Networks for
  Gravitational-Wave Inference}.
\newblock {\em \prl\/} {\bf 124} (2020) 041102.
\newblock \doi{10.1103/PhysRevLett.124.041102}.

\bibitem[{{Green} et~al.(2020){Green}, {Simpson}, and {Gair}}]{Green:20}
{Green} SR, {Simpson} C, {Gair} J.
\newblock {Gravitational-wave parameter estimation with autoregressive neural
  network flows}.
\newblock {\em \prd\/} {\bf 102} (2020) 104057.
\newblock \doi{10.1103/PhysRevD.102.104057}.

\bibitem[{{Talbot} and {Thrane}(2020)}]{Talbot:20}
{Talbot} C, {Thrane} E.
\newblock {Fast, flexible, and accurate evaluation of gravitational-wave
  Malmquist bias with machine learning}.
\newblock {\em arXiv e-prints\/}  (2020) arXiv:2012.01317.

\bibitem[{{Chatterjee} et~al.(2020){Chatterjee}, {Ghosh}, {Brady}, {Kapadia},
  {Miller}, {Nissanke} et~al.}]{Chatterjee:20}
{Chatterjee} D, {Ghosh} S, {Brady} PR, {Kapadia} SJ, {Miller} AL, {Nissanke} S,
  et~al.
\newblock {A Machine Learning-based Source Property Inference for Compact
  Binary Mergers}.
\newblock {\em \apj\/} {\bf 896} (2020) 54.
\newblock \doi{10.3847/1538-4357/ab8dbe}.

\bibitem[{{D'Emilio} et~al.(2021){D'Emilio}, {Green}, and
  {Raymond}}]{DEmilio:21}
{D'Emilio} V, {Green} R, {Raymond} V.
\newblock {Density estimation with Gaussian processes for gravitational wave
  posteriors}.
\newblock {\em \mnras\/} {\bf 508} (2021) 2090--2097.
\newblock \doi{10.1093/mnras/stab2623}.

\bibitem[{{{\'A}lvares} et~al.(2021){{\'A}lvares}, {Font}, {Freitas},
  {Freitas}, {Morais}, {Nunes} et~al.}]{Alvares:21}
{{\'A}lvares} JD, {Font} JA, {Freitas} FF, {Freitas} OG, {Morais} AP, {Nunes}
  S, et~al.
\newblock {Exploring gravitational-wave detection and parameter inference using
  deep learning methods}.
\newblock {\em Classical and Quantum Gravity\/} {\bf 38} (2021) 155010.
\newblock \doi{10.1088/1361-6382/ac0455}.

\bibitem[{{Williams} et~al.(2021){Williams}, {Veitch}, and
  {Messenger}}]{Williams:21}
{Williams} MJ, {Veitch} J, {Messenger} C.
\newblock {Nested sampling with normalizing flows for gravitational-wave
  inference}.
\newblock {\em \prd\/} {\bf 103} (2021) 103006.
\newblock \doi{10.1103/PhysRevD.103.103006}.

\bibitem[{{Krastev} et~al.(2021){Krastev}, {Gill}, {Villar}, and
  {Berger}}]{Krastev:21}
{Krastev} PG, {Gill} K, {Villar} VA, {Berger} E.
\newblock {Detection and parameter estimation of gravitational waves from
  binary neutron-star mergers in real LIGO data using deep learning}.
\newblock {\em Physics Letters B\/} {\bf 815} (2021) 136161.
\newblock \doi{10.1016/j.physletb.2021.136161}.

\bibitem[{{Xia} et~al.(2021){Xia}, {Shao}, {Zhao}, and {Cao}}]{Xia:21}
{Xia} H, {Shao} L, {Zhao} J, {Cao} Z.
\newblock {Improved deep learning techniques in gravitational-wave data
  analysis}.
\newblock {\em \prd\/} {\bf 103} (2021) 024040.
\newblock \doi{10.1103/PhysRevD.103.024040}.

\bibitem[{{Colgan} et~al.(2020){Colgan}, {Corley}, {Lau}, {Bartos}, {Wright},
  {M{\'a}rka} et~al.}]{Colgan:20}
{Colgan} RE, {Corley} KR, {Lau} Y, {Bartos} I, {Wright} JN, {M{\'a}rka} Z,
  et~al.
\newblock {Efficient gravitational-wave glitch identification from
  environmental data through machine learning}.
\newblock {\em \prd\/} {\bf 101} (2020) 102003.
\newblock \doi{10.1103/PhysRevD.101.102003}.

\bibitem[{{Essick} et~al.(2020){Essick}, {Godwin}, {Hanna}, {Blackburn}, and
  {Katsavounidis}}]{Essick:20}
{Essick} R, {Godwin} P, {Hanna} C, {Blackburn} L, {Katsavounidis} E.
\newblock {iDQ: Statistical Inference of Non-Gaussian Noise with Auxiliary
  Degrees of Freedom in Gravitational-Wave Detectors}.
\newblock {\em arXiv e-prints\/}  (2020) arXiv:2005.12761.

\bibitem[{{Cuoco} et~al.(2020){Cuoco}, {Powell}, {Cavagli{\`a}}, {Ackley},
  {Bejger}, {Chatterjee} et~al.}]{Cuoco:20}
{Cuoco} E, {Powell} J, {Cavagli{\`a}} M, {Ackley} K, {Bejger} M, {Chatterjee}
  C, et~al.
\newblock {Enhancing Gravitational-Wave Science with Machine Learning}.
\newblock {\em arXiv e-prints\/}  (2020) arXiv:2005.03745.

\bibitem[{{Torres-Forn{\'e}} et~al.(2020){Torres-Forn{\'e}}, {Cuoco}, {Font},
  and {Marquina}}]{Torres-Forne:20}
{Torres-Forn{\'e}} A, {Cuoco} E, {Font} JA, {Marquina} A.
\newblock {Application of dictionary learning to denoise LIGO's blip noise
  transients}.
\newblock {\em \prd\/} {\bf 102} (2020) 023011.
\newblock \doi{10.1103/PhysRevD.102.023011}.

\bibitem[{{Biswas} et~al.(2020){Biswas}, {McIver}, and {Mahabal}}]{Biswas:20}
{Biswas} A, {McIver} J, {Mahabal} A.
\newblock {New methods to assess and improve LIGO detector duty cycle}.
\newblock {\em Classical and Quantum Gravity\/} {\bf 37} (2020) 175008.
\newblock \doi{10.1088/1361-6382/ab8650}.

\bibitem[{{Soni} et~al.(2021){Soni}, {Berry}, {Coughlin}, {Harandi}, {Jackson},
  {Crowston} et~al.}]{Soni:21}
{Soni} S, {Berry} CPL, {Coughlin} SB, {Harandi} M, {Jackson} CB, {Crowston} K,
  et~al.
\newblock {Discovering features in gravitational-wave data through detector
  characterization, citizen science and machine learning}.
\newblock {\em Classical and Quantum Gravity\/} {\bf 38} (2021) 195016.
\newblock \doi{10.1088/1361-6382/ac1ccb}.

\bibitem[{{Sankarapandian} and {Kulis}(2021)}]{Sankarapandian:21}
{Sankarapandian} S, {Kulis} B.
\newblock {$\beta$-Annealed Variational Autoencoder for glitches}.
\newblock {\em arXiv e-prints\/}  (2021) arXiv:2107.10667.

\bibitem[{{Zhan} et~al.(2021){Zhan}, {Jia}, {Ma}, {Lu}, and {Lin}}]{Zhan:21}
{Zhan} C, {Jia} M, {Ma} C, {Lu} Z, {Lin} W.
\newblock {The response of the Convolutional Neural Network to the transient
  noise in Gravitational wave detection}.
\newblock {\em arXiv e-prints\/}  (2021) arXiv:2103.03557.

\bibitem[{Mogushi(2021)}]{Mogushi:21b}
Mogushi K.
\newblock Reduction of transient noise artifacts in gravitational-wave data
  using deep learning.
\newblock {\em arXiv preprint arXiv:2105.10522\/}  (2021).

\bibitem[{Chollet et~al.(2015)}]{Chollet:15}
[Dataset] Chollet F, et~al.
\newblock Keras.
\newblock \url{https://keras.io} (2015).

\bibitem[{Abadi et~al.(2015)Abadi, Agarwal, Barham, Brevdo, Chen, Citro
  et~al.}]{Abadi:15}
[Dataset] Abadi M, Agarwal A, Barham P, Brevdo E, Chen Z, Citro C, et~al.
\newblock {TensorFlow}: Large-scale machine learning on heterogeneous systems
  (2015).
\newblock Software available from tensorflow.org.

\bibitem[{{Barsotti} et~al.(2010){Barsotti}, {Evans}, and
  {Fritschel}}]{Barsotti:10}
{Barsotti} L, {Evans} M, {Fritschel} P.
\newblock {Alignment sensing and control in advanced LIGO}.
\newblock {\em Classical and Quantum Gravity\/} {\bf 27} (2010) 084026.
\newblock \doi{10.1088/0264-9381/27/8/084026}.

\bibitem[{Yu(2019)}]{Yu:19}
Yu H.
\newblock {\em {Astrophysical signatures of neutron stars in compact binaries
  and experimental improvements on gravitational-wave detectors}\/}.
\newblock Ph.D. thesis, MIT (2019).

\bibitem[{{Sidles} and {Sigg}(2006)}]{Sidles:06}
{Sidles} JA, {Sigg} D.
\newblock {Optical torques in suspended Fabry Perot interferometers}.
\newblock {\em Physics Letters A\/} {\bf 354} (2006) 167--172.
\newblock \doi{10.1016/j.physleta.2006.01.051}.

\bibitem[{{Hirose} et~al.(2010){Hirose}, {Kawabe}, {Sigg}, {Adhikari}, and
  {Saulson}}]{Hirose:10}
{Hirose} E, {Kawabe} K, {Sigg} D, {Adhikari} R, {Saulson} PR.
\newblock {Angular instability due to radiation pressure in the LIGO
  gravitational-wave detector}.
\newblock {\em \ao\/} {\bf 49} (2010) 3474.
\newblock \doi{10.1364/AO.49.003474}.

\bibitem[{{Dooley} et~al.(2013){Dooley}, {Barsotti}, {Adhikari}, {Evans},
  {Fricke}, {Fritschel} et~al.}]{Dooley:13}
{Dooley} KL, {Barsotti} L, {Adhikari} RX, {Evans} M, {Fricke} TT, {Fritschel}
  P, et~al.
\newblock {Angular control of optical cavities in a
  radiation-pressure-dominated regime: the Enhanced LIGO case}.
\newblock {\em Journal of the Optical Society of America A\/} {\bf 30} (2013)
  2618.
\newblock \doi{10.1364/JOSAA.30.002618}.

\bibitem[{Black et~al.(2010)Black, Chelermsongsak, DeSalvo, Korth, Barton, Cook
  et~al.}]{Black:10}
Black ED, Chelermsongsak T, DeSalvo R, Korth Z, Barton M, Cook D, et~al.
\newblock {\em {Advanced-LIGO Optical Levers Design Requirements}\/} (2010).
\newblock {L}IGO Document T0900174.

\bibitem[{{George} and {Huerta}(2018{\natexlab{a}})}]{George:18a}
{George} D, {Huerta} EA.
\newblock {Deep neural networks to enable real-time multimessenger
  astrophysics}.
\newblock {\em \prd\/} {\bf 97} (2018{\natexlab{a}}) 044039.
\newblock \doi{10.1103/PhysRevD.97.044039}.

\bibitem[{{George} and {Huerta}(2018{\natexlab{b}})}]{George:18b}
{George} D, {Huerta} EA.
\newblock {Deep Learning for real-time gravitational wave detection and
  parameter estimation: Results with Advanced LIGO data}.
\newblock {\em Physics Letters B\/} {\bf 778} (2018{\natexlab{b}}) 64--70.
\newblock \doi{10.1016/j.physletb.2017.12.053}.

\bibitem[{Yu and Adhikari(2021)}]{Yu:21z}
Yu H, Adhikari R.
\newblock {\em {Subtracting bilinear noise using machine-learning neural
  networks}\/} (2021).
\newblock {L}IGO Document G2100738.

\end{thebibliography}

\clearpage
\begin{table}[!h]
\caption{Network using a general structure. All the 20 auxiliary channels (16 slow channels and 4 fast ones) are directly input to the first layer and the nonlinearity is achieved by a series ELU activation in the hidden layers. In this network, there are about 863,000 trainable parameters in total. \label{tab:CNN_reg}}

\begin{tabular}{c|cccc}
&\textrm{layer}&
\textrm{output dimension}&\textrm{kernel size}&\textrm{activation} \\
\hline\hline
\multirow{11}{4em}{general structure} 
& Conv1D & 32 & 1024 & Linear  \\
& Dropout & -- & -- & rate=$8\times 10^{-6}$  \\
& Conv1D & 128 & 32 & Linear \\
& Conv1D & 16 & 8 & Linear \\
& Conv1D & 128 & 16 & Linear \\
& Dropout & -- & -- &  rate=$8\times 10^{-6}$ \\
& Dense & 128 & -- & ELU \\
& Dense & 64 & -- & ELU \\
& Dense & 32 & -- & ELU \\
& Dense & 8 & -- & ELU \\
& Dense & 1 & -- & Linear 
\end{tabular}
\end{table}

\begin{table}[!h]
\caption{Network explicitly utilizing the ``slow $\times$ fast structure''. We input the 16 slow channels to the slow path and use a series of linear operations (signal filtering and blending) to form 4 super-sensors that effectively monitors the spot position on each test masses. These super-sensors' output are then multiplied with the 4 angular witnesses input to the fast path by a multiply layer to eventually become a length signal. Note that all the activations are linear as the only nonlinearity is explicitly incorporated by the multiply layer. This network has about 736,000 trainable parameters in total. \label{tab:CNN_sxf}}
\begin{tabular}{c|cccc}
&\textrm{layer}&
\textrm{output dimension}&\textrm{kernel size}&\textrm{activation} \\
\hline\hline
\multirow{10}{4em}{slow path} 
& Conv1D & 32 & 1024 & Linear  \\
& Dropout & -- & -- & rate=$1\times 10^{-8}$  \\
& Conv1D & 128 & 32 & Linear \\
& Conv1D & 16 & 8 & Linear \\
& Conv1D & 128 & 16 & Linear \\
& Dropout & -- & -- &  rate=$1\times 10^{-8}$ \\
& Dense & 128 & -- & Linear \\
& Dense & 64 & -- & Linear \\
& Dense & 32 & -- & Linear \\
& Dense & 4 & -- & Linear \\
\hline
\multirow{4}{4em}{fast path} 
& Conv1D & 16 & 8 & Linear\\
& Dense & 64 & -- & Linear \\
& Dense & 32 & -- & Linear \\
& Dense & 4 & -- & Linear \\
\hline
\multirow{4}{4em}{after the multiply layer}
& Dense & 32 & -- & Linear \\
& Dense & 8 & -- & Linear \\
& Dense & 4 & -- & Linear \\
& Dense & 1 & -- & Linear 
\end{tabular}
\end{table}

\end{document}